\PassOptionsToPackage{unicode}{hyperref}
\PassOptionsToPackage{hyphens}{url}
\documentclass[
  11pt,
]{article}
\usepackage{lmodern, bm}
\usepackage{amssymb,amsmath}
\usepackage{ifxetex,ifluatex}
\ifnum 0\ifxetex 1\fi\ifluatex 1\fi=0 % if pdftex
  \usepackage[T1]{fontenc}
  \usepackage[utf8]{inputenc}
  \usepackage{textcomp} % provide euro and other symbols
\else % if luatex or xetex
  \usepackage{unicode-math}
  \defaultfontfeatures{Scale=MatchLowercase}
  \defaultfontfeatures[\rmfamily]{Ligatures=TeX,Scale=1}
\fi
\IfFileExists{upquote.sty}{\usepackage{upquote}}{}
\IfFileExists{microtype.sty}{% use microtype if available
  \usepackage[]{microtype}
  \UseMicrotypeSet[protrusion]{basicmath} % disable protrusion for tt fonts
}{}
\makeatletter
\@ifundefined{KOMAClassName}{% if non-KOMA class
  \IfFileExists{parskip.sty}{%
    \usepackage{parskip}
  }{% else
    \setlength{\parindent}{0pt}
    \setlength{\parskip}{6pt plus 2pt minus 1pt}}
}{% if KOMA class
  \KOMAoptions{parskip=half}}
\makeatother
\usepackage{xcolor}
\IfFileExists{xurl.sty}{\usepackage{xurl}}{} % add URL line breaks if available
\IfFileExists{bookmark.sty}{\usepackage{bookmark}}{\usepackage{hyperref}}
\hypersetup{
  pdftitle={A Bayesian cohort component projection model to estimate adult populations at the subnational level},
  pdfauthor={Monica Alexander; Leontine Alkema},
  hidelinks,
  pdfcreator={LaTeX via pandoc}}
\usepackage[margin=1in]{geometry}
\usepackage{graphicx,grffile}
\usepackage{rotating}
\usepackage{caption}
\usepackage{subcaption}
\makeatletter
\def\maxwidth{\ifdim\Gin@nat@width>\linewidth\linewidth\else\Gin@nat@width\fi}
\def\maxheight{\ifdim\Gin@nat@height>\textheight\textheight\else\Gin@nat@height\fi}
\makeatother
\setkeys{Gin}{width=\maxwidth,height=\maxheight,keepaspectratio}
\makeatletter
\def\fps@figure{htbp}
\makeatother
\usepackage{setspace}
\usepackage{booktabs}
\definecolor{OliveGreen}{RGB}{34,139,34}

\title{A Bayesian cohort component projection model to estimate adult
populations at the subnational level in data-sparse settings}
\author{Monica Alexander\footnote{University of Toronto.
  \texttt{monica.alexander@utoronto.ca}.} \and Leontine Alkema\footnote{University of Massachusetts, Amherst.
  \texttt{lalkema@umass.edu}. The work was supported by the Bill \& Melinda Gates Foundation. We thank Gregory Guranich for assistance with R programming. }}
\date{}

\begin{document}
\maketitle
\begin{abstract}
Accurate estimates of subnational populations are important for policy formulation and monitoring population health indicators. For example, estimates of the number of women of reproductive age are important to understand the population at risk to maternal mortality and unmet need for contraception. However, in many low-income countries, data on population counts and components of population change are limited, and so levels and trends subnationally are unclear. We present a Bayesian constrained cohort component model for the estimation and projection of subnational populations. The model builds on a cohort component projection framework, incorporates census data and estimates from the United Nation's World Population Prospects, and uses characteristic mortality schedules to obtain estimates of population counts and the components of population change, including internal migration. The data required as inputs to the model are minimal and available across a wide range of countries, including most low-income countries. The model is applied to estimate and project populations by county in Kenya for 1979-2019, and validated against the 2019 Kenyan census.
\end{abstract}

\newpage

\hypertarget{introduction}{%
\section{Introduction}\label{introduction}}

Reliable estimates of demographic and health indicators at the
subnational level are essential for monitoring trends and inequalities
over time. As part of monitoring progress towards global health targets
such as the Sustainable Development Goals (SDGs), there has been
increasing recognition of the substantial differences that can occur
across regions within a country (World Health Organization (WHO) 2016b;
Lim et al. 2016; He et al. 2017). As such, analysis of national-level
trends is often inadequate, and subnational patterns should be
considered in order to fully understand likely future trajectories.
Indeed, estimates and projections of important indicators such as child
mortality and contraceptive use are now being published at the
subnational level (New et al. 2017; Wakefield et al. 2019).

To effectively measure health indicators of interest, we need to be able
to accurately estimate the size of the population at risk. In order to
convert the \emph{rate} of incidence of a particular demographic or
health outcome to the \emph{number} of people affected by that outcome,
we need a good estimate of the denominator of those rates. As such,
population counts are essential knowledge for policy planning and
resource allocation purposes. However, even something as seemingly
simple as the number of people in an area of a certain age is relatively
unknown in many countries, particularly low-income countries that do not
have well-functioning vital registration systems.
And as previously reported outcomes show, differences in estimates of
the population at risk can have a large effect on the resulting
estimates of key indicators. For example, in 2017 the United Nations
Inter-agency Group for Child Mortality Estimation (UN-IGME) and the
Institute for Health Metrics and Evaluation (IHME) both published
estimates of under-five child mortality in countries worldwide (GBD 2016
Mortality Collaborators (IHME) 2017; UN-IGME 2017). However, estimates
for 2016 differed markedly, with IHME's estimate being 642,000 deaths
lower than the UN-IGME estimate. The main reason for the discrepancy was
the different sets of estimates of live births: IHME assumed there were
128.8 million live births in 2016, which was 12.2 million lower than the
141 million used by UN-IGME.

Data on population counts by age and sex at the subnational level vary
substantially by country, and often data availability and quality is the
worst in countries where outcomes are also relatively poor. For example,
many low-income countries may only have one or two historical censuses
available, and very little data available on internal migration or
mortality rates at the subnational level. This situation is in stark
contrast to many high-income countries where multiple data sources on
population counts, mortality and migration may exist. These varying data
availability contexts both present challenges in estimates of population
and the components of population change. In data-rich contexts, the
challenge is to reconcile multiple data sources that may be measuring
the same outcome. In data-sparse contexts, the challenge is to obtain
reasonable estimates without many observations. In both cases
traditional demographic models are often utilized, which often center
around a cohort component projection framework and take advantage of the
fact that patterns in populations often exhibit strong regularities
across age and time. However, these classical methods do not give any
indication of uncertainty around the estimates or projections, and
incorporating information from different data sources often requires
adhoc adjustments to ensure consistency. To overcome these limitations,
we propose a method that builds on classical demographic estimation of
populations by incorporating these techniques within a probabilistic
framework.

In particular, we present a Bayesian constrained cohort component model
to estimate subnational adult populations, focusing on women of
reproductive age (WRA), i.e.~women aged 15-49. This subgroup forms the
population at risk for many important health indicators such as
fertility rates, maternal mortality, and measures of contraceptive
prevalence. The model presented embeds a cohort component projection
setup in a Bayesian framework, allowing uncertainty in data and
population processes to be taken into account. At a minimum, the model
uses data on population and migration counts from censuses, as well as
national-level information on mortality and population trends, taken
from the UN World Population Prospects (UNPD 2019a). As data
requirements are relatively small, the methodology is applicable across
a wide range of countries, and overcomes limitations of previous
subnational cohort component methods, which require relatively large
amounts of data. Estimates and projections of population by age are
produced, as well as estimates of subnational mortality schedules and
in- and out- migration flows. As such, results from the model help to
understand population at risk to demographic and health outcomes at the
subnational level, but also to understand drivers of population change
and how these may in turn affect trends in indicators of interest.

The remainder of this paper is structured as follows. The next section
gives a brief overview of existing methods of subnational population
estimation, and outlines the contributions of the model proposed here. 
We then describe the main data sources typically available for subnational population estimation in low-income countries, using counties in Kenya as an example, followed by detailed
description of the proposed methodology. We then present results of fitting the model to data in Kenya and validate its out-of-sample projections against the 2019 census. Finally, possible extensions are discussed.

\hypertarget{existing-methods-of-subnational-population-estimation}{%
\section{Existing methods of subnational population
estimation}\label{existing-methods-of-subnational-population-estimation}}

Methods to estimate population at the subnational level are similar to
estimation methods at the national level. However, there are several
notable challenges of subnational population estimation that do not
exist at a country level. Firstly, migration flows are more important at
the subnational level. While migration flows are often assumed to be
negligible at the national level, they are usually larger as a
proportion of total population size at the regional level. In addition,
migration flows at the subnational level are also often more difficult
to estimate. Any particular region could have net in- or out-migration,
and flows to and from different regions can differ markedly in
magnitude. Secondly, when estimating subnational populations, it is
important to ensure the sum of all regions agrees with national
estimates produced elsewhere. In practice, this usually involves a
process of calibration against a known national population so that they match the
total. Lastly, data quality and availability is often poorer at the
subnational level. Populations at the regional level are smaller and
data are often more volatile, and data on key indicators of mortality
and internal migration is often lacking or unreliable.

\hypertarget{traditional-methods}{%
\subsection{Traditional methods}\label{traditional-methods}}

Perhaps the simplest and least data-intensive methods of subnational
population estimation involve interpolation and extrapolation of
regional shares of the total population (Swanson and Tayman 2012). Given
two (or more) censuses, one can calculate the relevant shares of the
population by age, sex and region and see how they have changed over
time. Intercensal estimations of populations assume constant increase
(or decrease) over time. Projection of populations into the future can
then be made based on assumptions of constant levels or trends in
shares. For example, the U.S. census Bureau produce subnational
population estimates for the majority of countries worldwide (U.S.
census Bureau 2017). The methods used to produce such estimates involve
making assumptions such as constant or logistic growth, and iteratively
calculating population proportions by age, sex and region such that they
match the country's total populations (Leddy 2017).

The most commonly used methods of population estimation and projection
are cohort component methods. These center on the demographic accounting
identity, which states that the population size (\(P\)) at time \(t\) is
equal to the population size at \(t-1\), plus births (\(B\)) and
in-migrants (\(I\)), minus deaths (\(D\)) and out-migrants (\(O\))
(Wachter 2014): \begin{equation}
P_{t} = P_{t-1} + B_{t-1} + I_{t-1} - D_{t-1} - O_{t-1}
\end{equation} The above equation is for a total population, but the
same accounting equation holds for each age group separately (where
births only affect the first age group). The cohort component method of
population projection (Leslie 1945) takes a baseline population with a
certain age structure and survives it forward based on age-specific
mortality, fertility and migration rates. Cohort component methods are
important because they allow for overall population change to be related
to the main components of that change. By estimating population size
based on the components of fertility, mortality and migration, the
method allows changes in these components to be taken into account.
However, cohort component methods are more data-intensive than
extrapolation methods, which is particularly an issue at the subnational
level. For developing countries in particular, where well-functioning
vital registration systems do not exist, sufficient data on mortality,
fertility and migration is often lacking.

Other methods of subnational estimation involve building regression
models which relate other variables of interest to changes in population
over time. For example, one could regress the ratio of census
populations (area of interest / total population) against the ratio of
some other indicator e.g.~births, deaths, voters, school enrollments
(see Swanson and Tayman (2012) for a detailed review). However, given
the lack of data available in many developing countries -- on population
counts, let alone other indicators of growth -- these methods have
limited use in our context.

These traditional methods of population estimation are deterministic and
do not account for random variation in demographic processes and
possible measurement errors that may exist in the data. In practice, the
population data that are available in developing countries are often
sparse and may suffer from various types of errors. When estimating and
projecting population sizes through time, it is particularly important
in developing country contexts to give some indication of the level of
uncertainty around those estimates, based on stochastic error,
measurement error and uncertainties in the underlying modeling process.

\hypertarget{bayesian-methods}{%
\subsection{Bayesian methods}\label{bayesian-methods}}

The use of Bayesian methods in demography has become increasingly
common, as it provides a useful framework to incorporate different data
sources in the same model, account for various types of uncertainty, and
allow for information exchange across time and space (Bijak and Bryant
2016). Bayesian methods have been used to model and forecast national
populations (Raftery et al. 2012; UNPD 2019a), fertility (Alkema et al.
2011), mortality (Alexander and Alkema 2018; Alkema and New 2014; Girosi
and King 2008) and migration (Bijak 2008). In terms of estimating the
full demographic accounting identity, Wheldon et al. (2013) propose a
method for the reconstruction of past populations. The model embeds the
demographic accounting equation within a Bayesian hierarchical
framework, using information from available censuses to
reconstruct historical populations via a cohort component projection
framework. The authors show the method works well to estimate
populations and quantify uncertainty in a wide range of countries with
varying data availability (Wheldon et al. 2016). The method presented in
Wheldon et al. (2013) is designed for population reconstruction at the
national level, and as such, accounting for internal migration is not an
issue. In addition, their method relies upon and calibrates to national
population estimates produced as part of the UN World Population
Prospects.

In the field of subnational estimation, Bayesian methods have also been
used in many different contexts. For subnational mortality estimation,
many researchers have used Bayesian hierarchical frameworks to share
information about mortality trends across space and time, in contexts
where the available data are both reliable (Congdon, Shouls, and Curtis
1997; Alexander, Zagheni, and Barbieri 2017) and sparse (Schmertmann and
Gonzaga 2018). For subnational fertility estimation, Sevcikova, Raftery,
and Gerland (2018) propose a Bayesian model that produces estimates and
projections of subnational total fertility rates (TFRs) that are
consistent with national estimates of TFR produced by the UN. Building
from the local level up, Schmertmann et al. (2013) propose a method
which uses empirical Bayesian methods to smooth volatile fertility data
at the regional level, before modeling using a Brass relational model
variant.

In terms of population estimation at the subnational level, John Bryant
and colleagues have shown how the demographic accounting equation can be
placed within a Bayesian framework to account for and reconcile
different data sources population counts and the components of
population change (Bryant and Graham 2013; Bryant and Zhang 2018).
Bryant and Zhang (2018) show how the underlying demographic processes
can be captured through a process or system model, and different types
of uncertainty around data inputs is captured through data models. The
focus of Bryant and Graham (2013) is producing subnational population
estimates for New Zealand, reconciling and incorporating information
about the population from sources such as censuses, and school and
voting enrollments. The approach that we take in this paper is similar
to the Byrant et al.~approach, in that we model population change with a
process model, the components of which are described by system models,
and different sources of information are combined through the use of
data models. However, whereas Bryant et al.~tries to overcome challenges
of combining multiple data sources that may be measuring the same
outcome, we are trying to overcome the challenges of estimating
subnational populations in contexts where there is extremely limited
amounts of data available.

There is an increasing amount of work using geo-located data and
satellite imagery to estimate population sizes and flows in developing
countries (Wardrop et al., 2019, Leasure
et al. 2020). Led by the WorldPop project at the
University of Southampton (WorldPop 2018), researchers have used
information from satellite imagery to identify areas of settlements, and
combined this information with census data to obtain highly granular
population density estimates across Africa (Linard et al. 2012; Leasure
et al. 2020). While this work contributes to information about subnational populations, % at a fine-grained resolution of this work has its
%advantages, 
the focus and goals of this estimation work are different to
our goals in this paper. In particular, the goal of the WorldPop work is
primarily to obtain estimates of total population and population density
at a very granular level, rather than obtaining population estimates by
age and sex. The results have then been combined with data on age- and
sex-distributions from censuses (or more recent surveys) to map the
distribution of populations by age and sex. However, little attention is
paid to how age distributions across regions change over time. But
changes in age distributions are important in understanding broader
population change and how this will impact global health indicators of
interest. In addition, our approach is grounded in understanding the
main components of demographic change -- mortality and migration -- over time and how they affect population sizes, rather than just estimating the population size as a
single outcome.

The methodology proposed in this paper incorporates a cohort component
projection model into a Bayesian hierarchical framework to understand
changes in population structures over time. It allows estimates to be
driven by available data and for uncertainty to be incorporated around
estimates and projections. The approach has similarities with
methodologies described in Wheldon et al. (2013) (but with a focus on
subnational estimation) and in Bryant et al (2013; 2018) (but with a
focus on data-sparse situations). 

In particular, we introduce a framework to estimate subnational population counts and components of population change that relies on a minimal amount of data that is available for the vast majority of countries worldwide. Observations on subnational population counts an internal migration movements are taken from censuses, but no information on subnational mortality patterns is required. We instead use a mortality model approach based on principal components derived from national mortality schedules. Using principal components for demographic modeling and forecasting first gained popularity after Lee and Carter used the
technique as a basis for forecasting US mortality rates (Lee and Carter
1992). More recently, principal components has become increasingly used in demographic
modeling, in both fertility and mortality settings (Schmertmann et al.
2014; Clark 2016; Alexander, Zagheni, and Barbieri 2017).

While one strength of our approach is being able to estimate components of subnational population change with limited data, another strength of the proposed framework is that it can be readily extended to include other data or estimates. For example, gridded estimates produced as part of the WorldPop project could conceivably be treated as an additional data input to the model. 

\hypertarget{data}{%
\section{Data}\label{data}}

We aim to estimate female population counts for ages 15-49 per 5-year age group for
subnational areas that are the second administrative level down. This
data description focuses on Kenya, for which the model is applied in
later sections. However, the data and methods are more broadly
applicable to other countries that have similar data available. Inputs
used to obtain estimates come from two main sources: micro-level data
from censuses, and national population and mortality estimates from the
2019 World Population Prospects. These data sources are outlined in the
following sections.

\hypertarget{overview-of-kenyan-geography}{%
\subsection{Overview of Kenyan
geography}\label{overview-of-kenyan-geography}}

In Kenya, the first administrative units are provinces, and the second
administrative units are counties. There are eight provinces, including
the capital Nairobi, and 47 counties. The county boundaries have changed
over time, but have been stable since the 2009 census. We aim to produce
estimates of populations of women of reproductive age at the county
level based on county boundaries in 2009. Within the model, we also make
use of harmonized district boundaries (see description below), which are
slightly larger than counties. There are a total of 35 districts.
Provinces and districts are illustrated below in Fig \ref{map}.

\begin{figure}[h!]
\centering
\includegraphics[width=0.9\textwidth,height=\textheight]{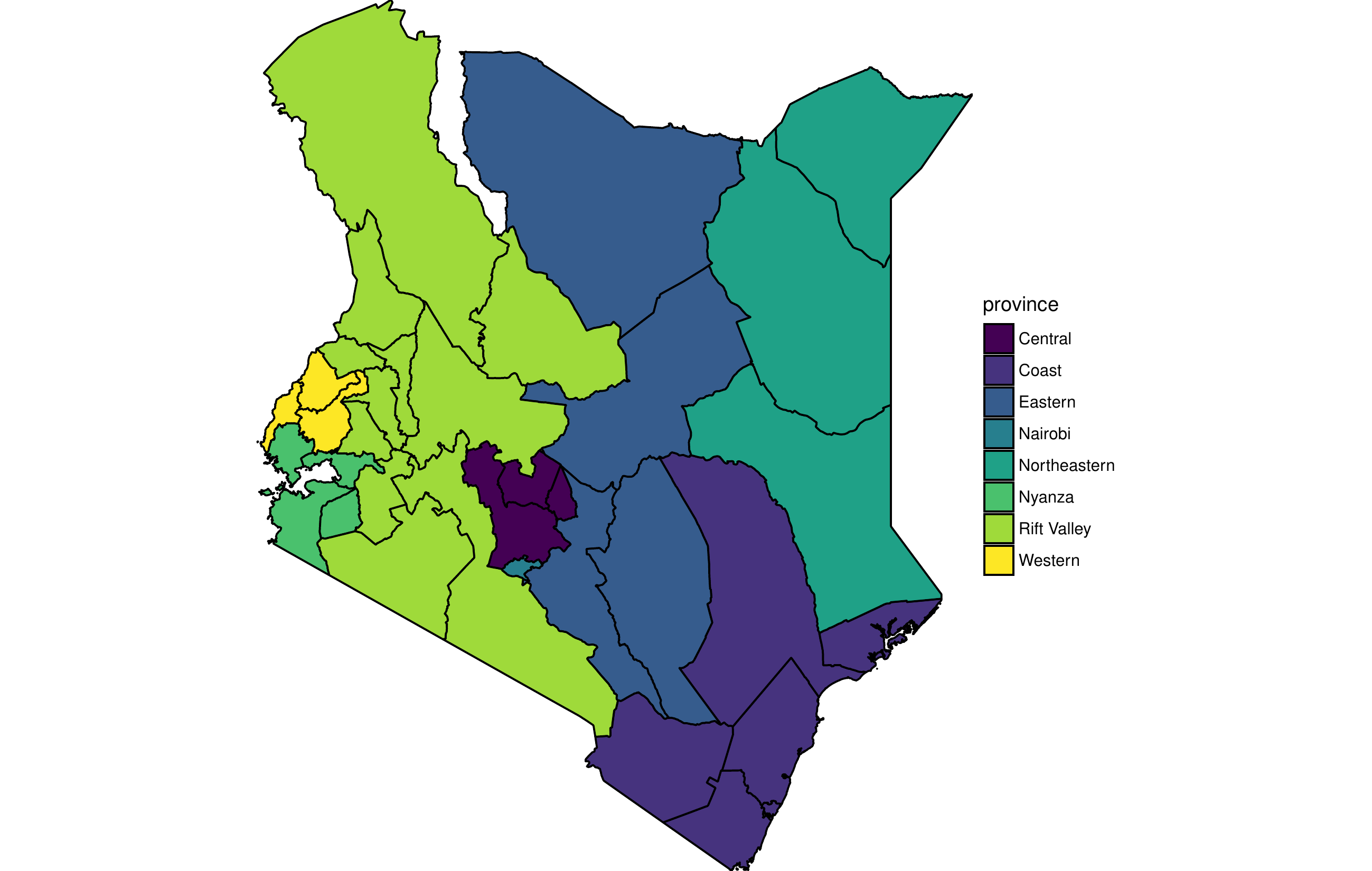}
\caption{Map of Kenya provinces, showing IPUMS harmonized districts.
\label{map}}
\end{figure}

\hypertarget{census-data}{%
\subsection{Census data}\label{census-data}}

Data inputs on subnational population counts and internal migration
flows come from national censuses. The census data are available through
Integrated Public Use Microdata Series (IPUMS) International (Minnesota
Population Center 2017). IPUMS-International contains samples of
microdata for 305 censuses over 85 different countries. The majority of
countries of interest have relatively recent censuses available through
IPUMS-International. Kenya has decennial censuses available from 1979 to
2009. Micro-level data are not available for the 2019 Kenyan census,
although population counts by sex and five-year age group and county are
available through the national statistics office. The 2019 data are
reserved for model evaluation, as detailed in the Model Evaluation
section.

In the micro-level IPUMS data, location of residence is reported at the
first (province) and second (county) administrative levels, as well as a
harmonized district level. For Kenya, the provinces are stable over
time, but before 2009 the county boundaries changed. As such, we only
have data at the county level for Kenya for 2009. However, we can make
use of the harmonized districts for data in years prior to 2009. The
districts represent slightly larger groups than the 47 Kenyan counties,
which are harmonized and temporally stable (IPUMS 2018). In all cases,
each 2009 county is completely contained in one unique district.

We used census data to obtain information on two different quantities:
observed population counts; and observed patterns of in- and
out-migration. Female population counts by five-year age
groups for ages 15-49 and subnational administrative region are obtained
directly from the IPUMS-International microdata. As these data are
samples (most commonly 10\%), the microdata are multiplied by the person
weights to obtain counts by age and
area.\footnote{The sampling error introduced by considering sampled microdata is accounted for in the data model, refer to the Methods section for details.}

Information on internal migration between counties and districts is also
obtained from national censuses. This is based on questions about a migrant's location of residence one year ago. We calculated in- and out-migration counts by age group, for
each region and each census. For the 2009 census, the calculations were
based on counties; for all earlier years the calculations were based on
districts.

%For Kenya, migration counts can be derived using information from two questions:

% \begin{itemize}
% \tightlist
% \item
%   A person's location one year ago (IPUMS harmonized variable
%   \texttt{MIGRATE1}). The information is provided in the form of
%   categories, i.e.~same major administrative unit, different minor
%   administrative unit; different major and minor administrative unit;
%   abroad. Migrants are limited to those who were in a different minor or
%   major administrative unit one year ago.
% \item
%   A person's district of residence one year ago (for Kenya, this is the
%   variable \texttt{MIGKE}).
% \end{itemize}

% Given these two variables, we can calculate the migration by age group
% for a particular county/district in the year before the census. For
% example, to get counts for migration for Nairobi in 2008, use
% information from the 2009 census and:

% \begin{itemize}
% \tightlist
% \item
%   calculate the number of in-migrants by summing residents in Nairobi
%   who indicated they were in a different administrative unit one year
%   ago;
% \item
%   calculate the number of out-migrants by summing residents in other
%   districts who indicated they lived in Nairobi one year ago
% \end{itemize}

\hypertarget{national-estimates-from-wpp}{%
\subsection{National estimates from
WPP}\label{national-estimates-from-wpp}}

The World Population Prospects (WPP) are the official population
estimates and projections produced by the United Nations. WPP is revised
every two years, with the latest revision being in 2019 (UNPD 2019a).
WPP estimates are produced using a combination of census and survey
data, and demographic and statistical methods. Both population counts
and mortality estimates from WPP are used in the model.

We use estimates from WPP 2019 in two ways. Firstly, we would like to
ensure that the sum of population estimates at the regional level agrees
with published estimates at the national level. National population
counts produced by WPP are used as a constraint in the model, subject to uncertainty. The WPP
models populations of five-year age groups every five years from
1950-2100.

National mortality estimates produced by WPP are used as the basis of a
mortality model for patterns at the regional level, capturing HIV/AIDS related patterns of mortality. WPP uses the
relationship between infant mortality and the probability of dying
between ages 15 and 60, i.e.~\(_{45}q_{15}\), to estimate a life table
based on Coale-Demeny Model Life Tables (UNPD 2019b). We use estimates
of the probability of dying between ages \(x\) and \(x+5\), \(_5q_x\).

\hypertarget{other-potential-data-sources}{%
\subsection{Other potential data
sources}\label{other-potential-data-sources}}

We use census data and WPP estimates as inputs to the model. There are
other available data sources that could be used as inputs. These sources
and the reasons for not including them are discussed in Appendix
\ref{other_data}.

\hypertarget{model}{%
\section{Model}\label{model}}

\subsection{Overview}

In this section we describe the modeling framework to estimate female
populations by five-year age group and county. The model is outlined for
the situation where, like in the Kenyan case, we do not observe
county-level information for every census, but we have information on
larger, harmonized districts that fully encapsulate the counties. This
situation is common for many low-income countries where geographic
boundaries may vary over time but there exist some other stably-defined
boundaries through the micro-data on IPUMS.

\begin{sidewaysfigure}
\includegraphics[width = \textwidth]{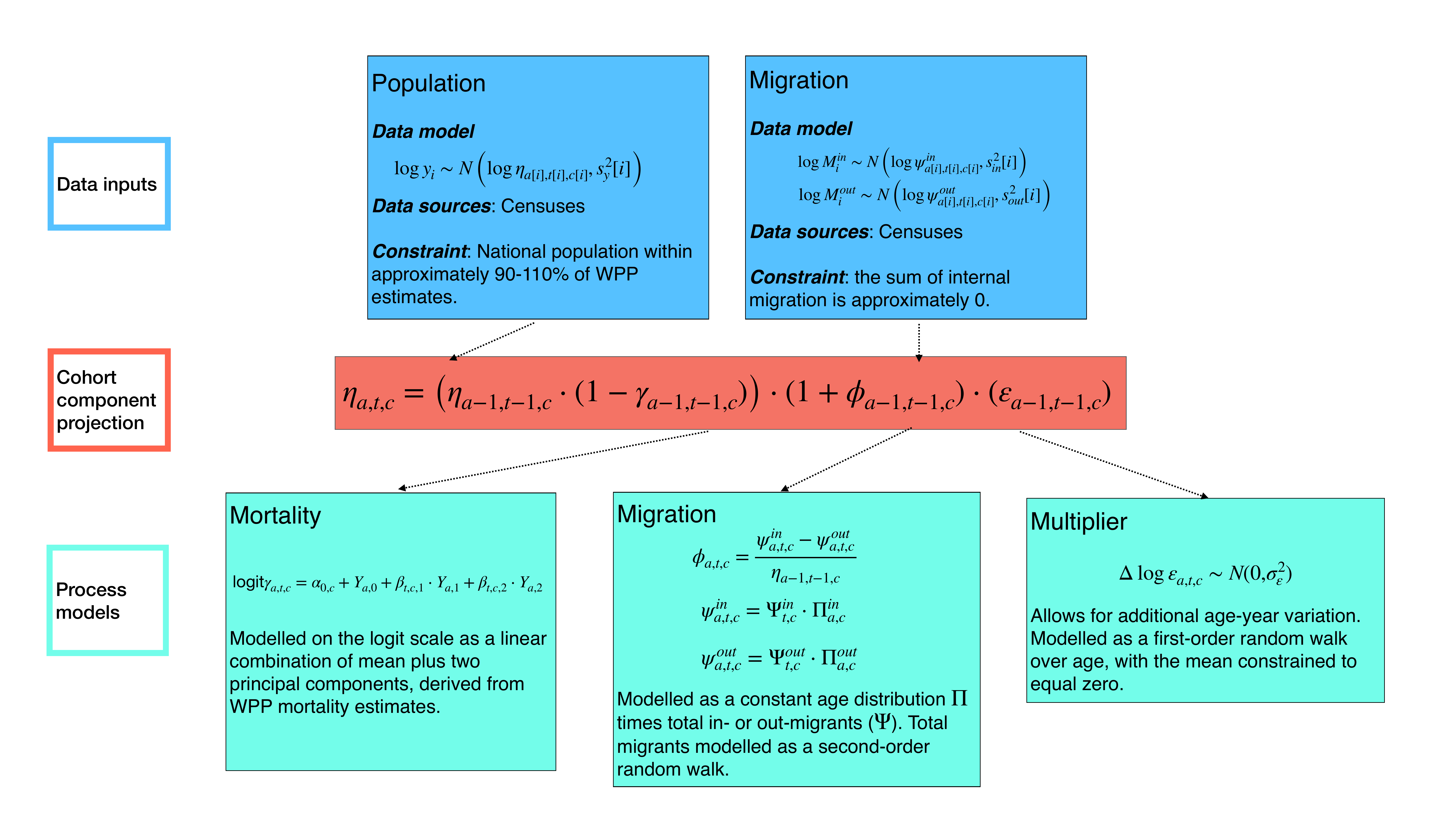}
\caption{Diagram showing the main components of the Bayesian cohort component projection model.}
\label{fig_diagram}
\end{sidewaysfigure}

There are many components and several types of data going into the model at
different stages. The overall model framework is summarized
visually in Figure \ref{fig_diagram}. We define \(\eta_{a,t,c}\) to be the underlying `true' population of women in
age group \(a\), year \(t\) and county \(c\). Our main modeling goal is
to obtain estimates and projections of these quantities. The population counts follow a cohort component projection (CCP) model, which assumes population counts in the current time period are those from the previous period, after taking into account expected changes in mortality and migration. The CCP model also includes an additional age-time multiplier which captures any other variation not already captured by expected changes in mortality or migration. Our set-up allows for changes in mortality and migration to be projected forward even if there are no data on these components, and is useful in data-sparse contexts where there is limited information available on the individual components of population change.

As illustrated in Figure \ref{fig_diagram}, the mortality, migration and additional age-time specific multipliers have additional `process models' (shown on the third row), and data on population counts and migration are related to the underlying process through data models (shown in the top row). 

The following sections broadly describe each component of the model. The full model specification and details can be found in Appendix \ref{appendix_model}. 

\hypertarget{population-model}{%
\subsection{Population model}\label{population-model}}

The model for
population includes: the cohort component project model, the data model, which relates observations of
population counts to the underlying quantities of interest; and the
national-level constraint, which ensures the sum of the county-level
populations is close to pre-published national estimates.

\hypertarget{ccp-model}{%
\subsubsection{Cohort component projection model}\label{ccp-model}}

The underlying population \(\eta_{a,t,c}\) can be expressed as
\begin{equation}
\label{eqn_ccp}
\eta_{a,t,c}= \left(\eta_{a-1,t-1,c} \cdot (1- \gamma_{a-1,t-1,c})\right) \cdot (1 + \phi_{a-1,t-1,c}) \cdot(\varepsilon_{a-1,t-1,c}),
\end{equation} where \(\gamma_{a,t,c}\) is the expected conditional probability
of death in age group \(a\), year \(t\) and county \(c\),
\(\phi_{a,t,c}\) is expected net migration (that is, in- minus out-migration) as
a proportion of population size, and \(\varepsilon_{a,t,c}\) is an
additional age-year-county multiplier. Note that this is a form of a
cohort component projection framework. As mentioned previously, our main
modeling goal is to obtain estimates of the \(\eta_{a,t,c}\), but we are
also interested in estimates of expected mortality (\(\gamma_{a,t,c}\)) and
expected migration (\(\phi_{a,t,c}\)), and, if non-zero, the multipliers ($\varepsilon_{a,t,c}$).

\hypertarget{data-model}{%
\subsubsection{Data model}\label{data-model}}

Define \(y_i\) to be \(i\)th observed population count. Depending on the
year of the census, \(y_i\) is either observed at the county \(c\) level
or district \(d\) level. The data model is: \begin{eqnarray}
    \log y_i|\eta_{a,t,c} &\sim& \begin{cases}
    N\left(\log \eta_{a[i],t[i],c[i]}, s^2_{y}[i]\right) \text{ if } t= 2009,\\
    N\left(\log \sum_{c \in d[i]}(\eta_{a[i],t[i],c[i]}), s^2_{y}[i]\right) \text{ if } t<2009,
    \end{cases}
\end{eqnarray} where \(s^2_y\) is the sampling error based on the fact
that the micro-data in IPUMS is a 10\% sample. The second case of the
above equation dictates that if we have observations prior to 2009, we
can only relate these to \(\eta_{a,t,c}\)'s that have been summed to the
district level.

\hypertarget{constraints-on-national-population}{%
\subsubsection{Constraints on national
population}\label{constraints-on-national-population}}

We would like to ensure the county-level populations \(\eta_{a,t,c}\) imply
a national-level population that is consistent with previously-published
estimates in WPP. To do this, we implement the following constraint in
the model, which roughly corresponds to the sum of the subnational
populations in any age and year being within 90-110\% of WPP. Further details on the constraint and priors in the population model are given in Appendix \ref{appendix_model}.

\hypertarget{mortality-model}{%
\subsection{Mortality model}\label{mortality-model}}

Equation \ref{eqn_ccp} requires estimates of the expected conditional probability
of death in each age group, year and county. As discussed in the Data
section and appendix, we do not have reliable information about mortality by age at
the county level, and as such we use information about mortality trends
at the national level as the basis for a mortality model at the
subnational level. A semi-parametric model is used to capture the shape of
national mortality through age and time, while allowing for differences
by county. In particular, we model county mortality on the logit scale
as \begin{eqnarray}
    \text{logit} (\gamma_{a,t,c}) &=& \alpha_{0,c} + Y_{a,0} + \beta_{t,c,1}\cdot Y_{a,1} + \beta_{t,c,2}\cdot Y_{a,2},
\end{eqnarray} where \(Y_{a,0}\) is the mean age-specific logit
mortality schedule of the national mortality curves and \(Y_{,1}\) and
\(Y_{,2}\) are the first two principal components derived from
national-level mortality schedules. Modeling on the logit scale ensures
the death probabilities are between zero and one.

Principal components create an underlying structure of the model in which regularities in age patterns of human mortality can be expressed.
Many different kinds of shapes of mortality curves can be expressed as a
combination of the components. Incorporating more than one principal
component allows for greater flexibility in the underlying shape of the
mortality age schedule. 

Principal components were obtained from a decomposition on a matrix
which contains a set of standard mortality curves. As discussed in the
Data section, we used national Kenyan life tables published in the World
Population Prospects 2019. In particular, let \(\bf{X}\) be a
\(N \times G\) matrix of logit mortality rates, where \(N\) is the
number of years and \(G\) is the number of age-groups. In this case, we
had \(N = 16\) years (estimates every 5 years from 1950 to 2025) of
\(G = 7\) age-groups (\(15-19, 20-24, \dots, 45-49\)). The SVD of
\(\bf X\) is \begin{equation}
\bf X = \bf{UDV'},
\end{equation} where \(\bf U\) is a \(N \times N\) matrix, \(\bf D\) is
a \(N \times G\) matrix and \(\bf V\) is a \(G \times G\) matrix. The
first two columns of \(\bf V\) (the first two right-singular values of
\(\bf X\)) are \(Y_{1:A,1}\) and \(Y_{1:A,2}\).

\begin{figure}[h!]
\centering
\includegraphics{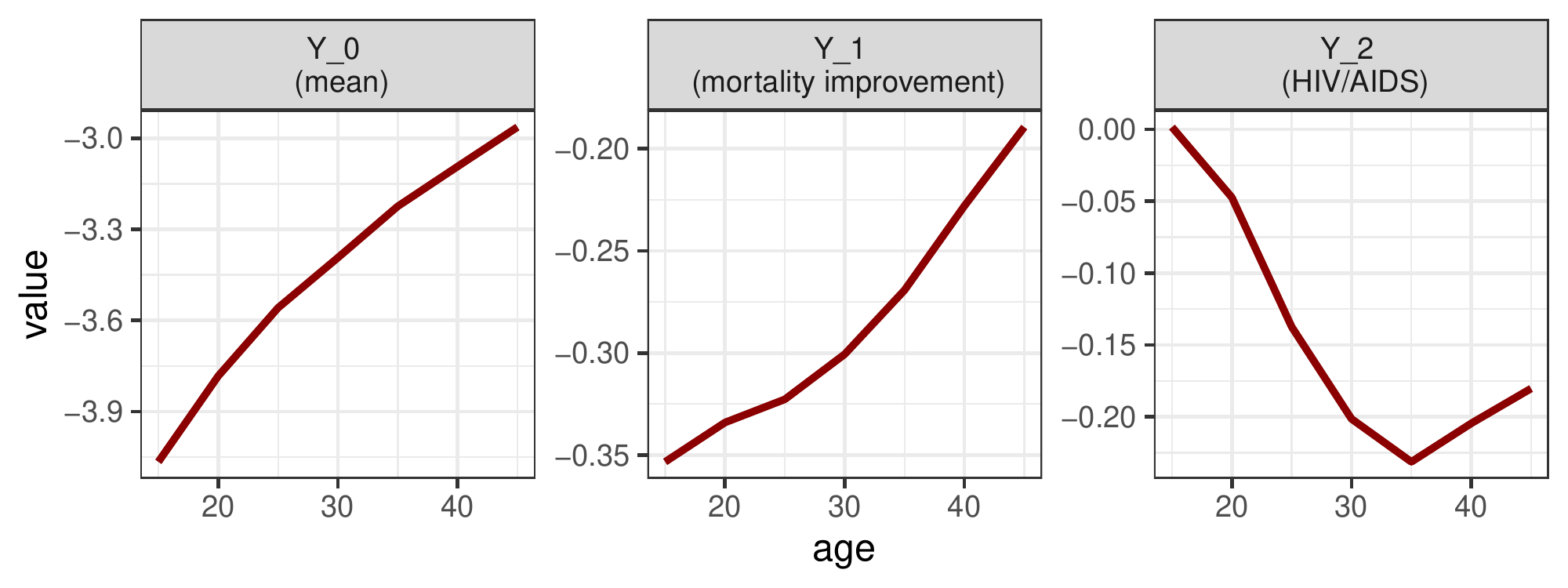}
\caption{Mean logit mortality schedule and first two principal
components. \label{pcs}}
\end{figure}

The mean mortality schedule and the first two principal components for
Kenyan national mortality curves between ages 15-49 from 1950--2020 are
shown in Fig.~\ref{pcs}. The mean logit mortality schedule shows a
standard age-specific mortality curve, with mortality increasing over
age. The first two principal components have demographic
interpretations. The first shows the average contribution of each age to
mortality improvement over time. This interpretation is similar to the
\(b_x\) term in a Lee-Carter model (Lee and Carter 1992). For the case
of Kenya, the second principal component most likely represents the
relative effect of HIV/AIDS mortality by age.

The county-specific mortality intercepts are modeled using a Normal
distribution centered at zero:

\begin{equation}
\alpha_{0,c}|\sigma^2_{\alpha} \sim N(0, \sigma^2_{\alpha}).
\end{equation}

The county-specific coefficients \(\beta_{t,c, k}\) are modeled as
fluctuations around a national mean: \begin{eqnarray}
    \beta_{t,c, k} &=& B^{nat}_{t,k} + \delta_{t,c, k},\\
    \delta_{t,c,k}|\delta_{t-1,c, k}, \sigma^2_{\delta} &\sim& N(\delta_{t-1,c, k},\sigma^2_{\delta}),
\end{eqnarray} where \(B^{nat}_{t,k}\) are the national coefficients on
principal components, derived from WPP data. The county-specific
fluctuations are modeled as a random walk.

\hypertarget{migration-model}{%
\subsection{Migration model}\label{migration-model}}

The second population change component of Equation \ref{eqn_ccp} refers to the
net-migration rate in a particular age group, year and county.
Specifically, define the net-migration rate as \begin{equation}
\phi_{a,t,c} = \frac{\psi_{a,t,c}^{in} - \psi_{a,t,c}^{out}}{\eta_{a-1,t-1,c}},
\end{equation} where \(\psi_{a,t,c}^{in}\) is the number of in-migrants and
\(\psi_{a,t,c}^{out}\) is the number of out-migrants.

For the migration component, we use observed data from the census. As such, in a similar way to the population model, we have a process
model, which defines the underlying migration process for the `true'
migrant parameters, and a data model, which relates observations from
the census to the underlying truth.

\hypertarget{process-model-1}{%
\subsubsection{Process model}\label{process-model-1}}

The model form for the number of in-migrants and out-migrants is
informed by patterns observed in the raw census data. In particular,
looking at the age distribution of both in- and out-migration (i.e.~the
proportion of total migrants who are in age group \(a\)) suggests that,
while the overall magnitude of migration changes over time, the age
patterns in migration are fairly constant (see figures in Appendix
\ref{mig_data}). This observation allowed us to simplify the expression
for the number of in-migrants and out-migrants, which are modeled as
\begin{eqnarray}
    \phi_{a,t,c} &=& \frac{\psi_{a,t,c}^{in} - \psi_{a,t,c}^{out}}{\eta_{a-1,t-1,c}},\\
    \psi_{a,t,c}^{in} &=& \Psi_{t,c}^{in}\cdot \Pi^{in}_{a,c},\\
    \psi_{a,t,c}^{out} &=& \Psi_{t,c}^{out}\cdot \Pi^{out}_{a,c},
\end{eqnarray} where \(\Psi_{t,c}^{in}\) and \(\Psi_{t,c}^{out}\) are the total number
of in- and out-migrants, respectively, and \(\Pi^{in}_{a,c}\) and
\(\Pi^{out}_{a,c}\) are the relevant age distributions. In this way the
age distributions are assumed to be constant over time while the total
counts vary. We model the total counts as a second order random walk to
impose a certain level of smoothness in the counts over time. As the model is meant to capture internal migration flows in and out of
each county, it must be the case that the sum of all in-migration flows
must equal the sum of all out-migration flows. As such, we also constrain
the difference between the sum of all estimated in- and out-migration
flows to be close to zero. See Appendix \ref{appendix_model} for further details. 

\hypertarget{data-model-1}{%
\subsubsection{Data model}\label{data-model-1}}

Finally, we relate the observed age-specific in- and out-migration
counts in the censuses, denoted \(M_i^{in}\) and \(M_i^{out}\), respectively, to
the underlying true counts \(\psi_{a,t,c}^{in}\) and \(\psi_{a,t,c}^{out}\) through
the following data model: \begin{eqnarray}
    \log M_i^{in}|\psi^{in}_{a,t,c} &\sim& \begin{cases}
    N\left(\log \psi^{in}_{a[i],t[i],c[i]}, s^2_{in}[i]\right) \text{ if } t[i]= 2009\\
    N\left(\log \sum_{c \in d[i]}(\psi^{in}_{a[i], t[i],c[i]}), s^2_{in}[i]\right) \text{ if } t[i]<2009
    \end{cases}\\
       \log M_i^{out}|\psi^{out}_{a,t,c} &\sim& \begin{cases}
    N\left(\log \psi_{a[i],t[i],c[i]}^{out}, s^2_{out}[i]\right) \text{ if } t[i]= 2009,\\
    N\left(\log \sum_{c \in d[i]}(\psi_{a[i],t[i],c[i]}^{out}), s^2_{out}[i]\right) \text{ if } t[i]<2009.
    \end{cases}
\end{eqnarray} In a similar fashion to the data model for population,
data observed prior to 2009 can only be related to the migration counts
that have been summed to the district level. In addition, the
\(s^2_{in}\) and \(s^2_{out}\) are the sampling errors based on the fact that the
micro-data in IPUMS is a 10\% sample. 

\hypertarget{additional-age-time-multiplier-varepsilon_atc}{%
\subsection{\texorpdfstring{Additional age-time multiplier
\(\varepsilon_{a,t,c}\)}{Additional age-time multiplier \textbackslash varepsilon\_\{a,t,c\}}}\label{additional-age-time-multiplier-varepsilon_atc}}

In both the models for expected mortality and migration discussed above, constraints are imposed on the age-specific effects. In particular, the use of the SVD approach to model mortality results in mortality age patterns that are linear combinations of the mean schedule and the components of change (the \(Y\)'s). Additionally, the migration model
assumes a constant age pattern of migration over time with varying
magnitudes of in- and out-migration. We assume these forms in order to
greatly reduce the number of parameters that need to be estimated in
each model, such that reasonable estimates of mortality and migration
rates can still be obtained in data-sparse settings.

In order to allow for county-specific age- and time- variation that may
not have already been captured by other components, we introduced an
additional age-time multiplier \(\varepsilon_{a,t,c}\) in the population
cohort component model (see Equation \ref{eqn_ccp}). We model these multipliers
on the log scale, and to ensure identifiability we assume the mean of
the sum of the log multipliers over all age groups is zero. This constraint is implemented
through the re-parameterization: \begin{eqnarray}
\log \bm{\varepsilon}_{1:A,t,c} &=& \mathbf{D(DD')}^{-1} \bm{\zeta}_{1:(A-1),t,c}, \\
\bm{\zeta}_{a,t,c} &\sim& N(0, \sigma^2_{\zeta}),
\end{eqnarray} where \(\mathbf D\) is first-order difference matrix (with
\({D}_{i,i} = -1\), \({D}_{i,i+1}=1\), and \({D}_{i,j} = 0\) otherwise) such that $\zeta_{a,t,c} = \log {\varepsilon}_{a,t,c} - \log \bm{\varepsilon}_{a-1,t,c}$.

\hypertarget{computation}{%
\subsection{Computation}\label{computation}}

The model was fitted in a Bayesian framework using the statistical
software R. Samples were taken from the posterior distributions of the
parameters via a Markov Chain Monte Carlo (MCMC) algorithm. This was
performed using JAGS software (Plummer 2003). Standard diagnostic checks
using trace plots and the $\hat{R}$ diagnostic (Gelman et al.\ 2020) were used to check convergence.

Best estimates of all parameters of interest were taken to be the median
of the relevant posterior samples. The 95\% Bayesian credible intervals
were calculated by finding the 2.5\% and 97.5\% quantiles of the
posterior samples.

% \hypertarget{code}{%
% \subsection{Code}\label{code}}

% Code available at XX

\hypertarget{results}{%
\section{Results}\label{results}}
In this section we illustrate some key results of population counts, mortality and migration. Additional results are presented in Appendix \ref{appendix_results}.

\hypertarget{population-estimates-and-projections}{%
\subsection{Population estimates and
projections}\label{population-estimates-and-projections}}

\begin{figure}[h!]
\centering
\includegraphics{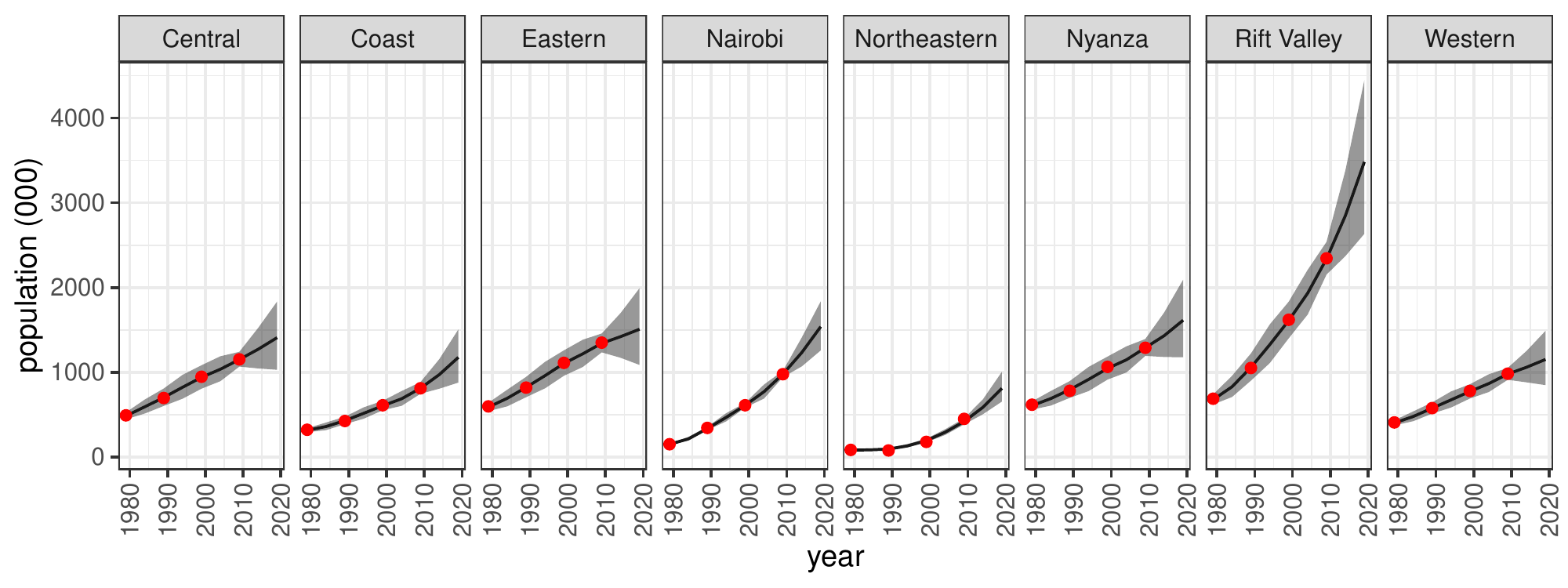}
\caption{Estimates of female population aged 15-49 by province, Kenya,
1979-2020. \label{province}}
\end{figure}

Figure~\ref{province} shows the WRA population by province in 1979-2019.
The black line and associated shaded area are the model estimates and
associated 95\% credible intervals. The red dots are the
data from decennial censuses. Populations
of WRA are increasing in every province, with the two largest provinces
being Nairobi and Rift Valley. While Northeastern is the smallest
province by population size, the growth rate is relatively rapid. This
is likely due to the relatively high fertility rates in this province
(Westoff and Cross 2006; Kenya National Bureau of Statistics 2015),
whereas rapid population increases in Nairobi are driven by
in-migration. %Fig.~\ref{district_map} shows population increases are concentrated in counties surrounding Nairobi.

%\begin{figure}[h!]
%\centering
%\includegraphics{fig/district_map.pdf}
%\caption{Estimates of female population aged 15-49 ('000) by county:
%1979, 1999 and 2019. \label{district_map}}
%\end{figure}

\begin{figure}[h!]
\centering
\includegraphics{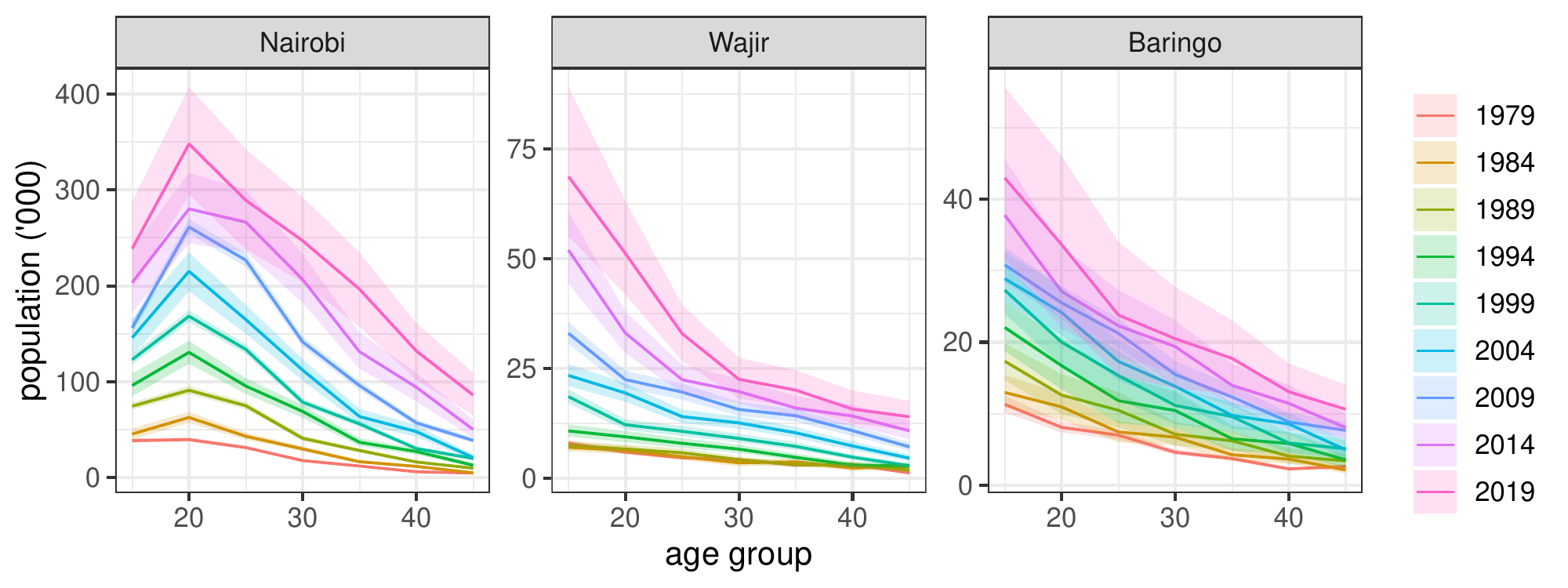}
\caption{Estimates of female population aged 15-49 ('000) by age and
year for three counties. \label{pop_age_year}}
\end{figure}

Figure \ref{pop_age_year} illustrates populations over age and time for
3 different counties. Note the different y-axis scales for each county. For Nairobi, populations are much larger and the presence of
net in-migration far surpasses the effects of mortality, leading to an
inverted-U shaped age distribution. For Wajir, a relatively rural county
in the northeast, population growth seems rapid over time. For Baringo, populations are relatively small and decline regularly over age due to
mortality. 

\hypertarget{mortality}{%
\subsection{Mortality}\label{mortality}}

\begin{figure}[h!]
\centering
\includegraphics{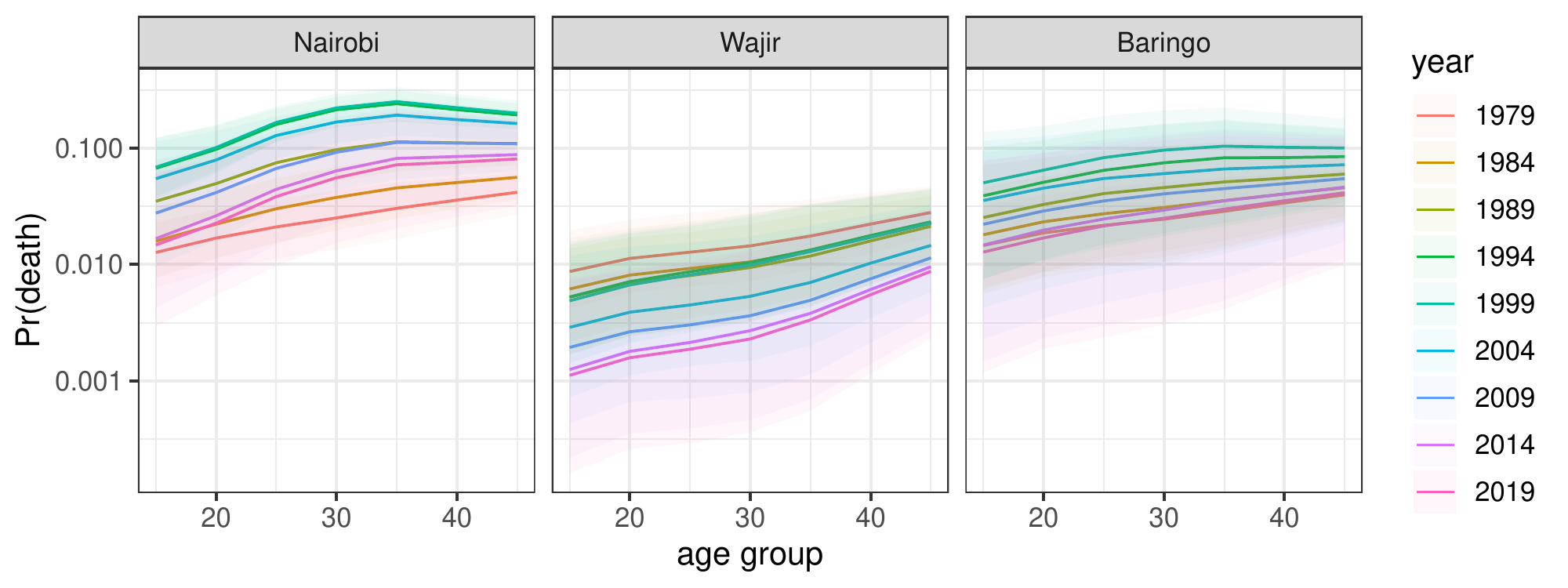}
\caption{Estimates of mortality by age and year for three counties.
\label{mort}}
\end{figure}

In addition to getting estimates of population counts, we also obtain
estimates of the components of population change, namely mortality and
migration. In terms of mortality, there is evidence of variation across
the counties. Focusing on the three counties as above, mortality profiles are quite different,
with Baringo's estimates being similar to the national mean (Figure
\ref{mort}).

% \begin{figure}[h!]
% \centering
% \includegraphics[width = 0.6\textwidth]{fig/mort_map.pdf}
% \caption{Estimates of relative level of overall mortality by district.
% \label{mortmap}}
% \end{figure}

\hypertarget{migration}{%
\subsection{Migration}\label{migration}}

\begin{figure}[h!]
\centering
\begin{subfigure}[b]{\textwidth}
\includegraphics[width=\textwidth]{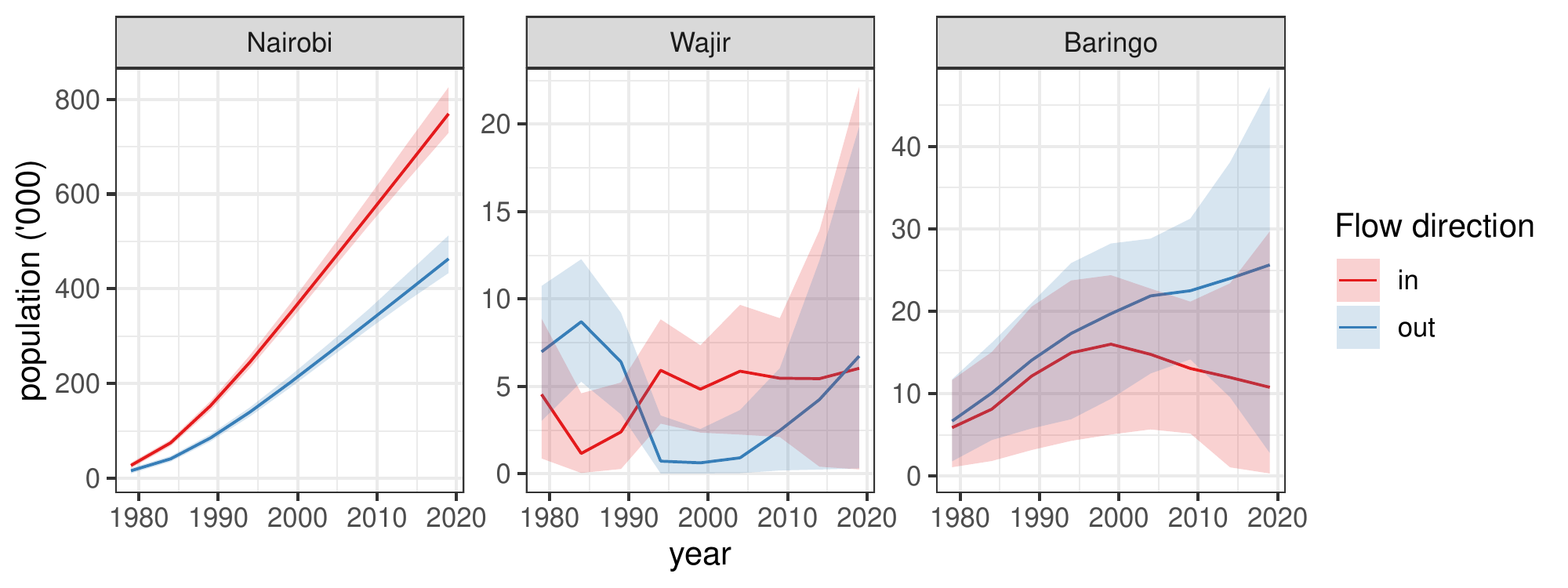}
\caption{Estimates of total in- and out-migration over time.\label{mig_both}}
\end{subfigure}
\hfill
\begin{subfigure}[b]{\textwidth}
\includegraphics[width=\textwidth]{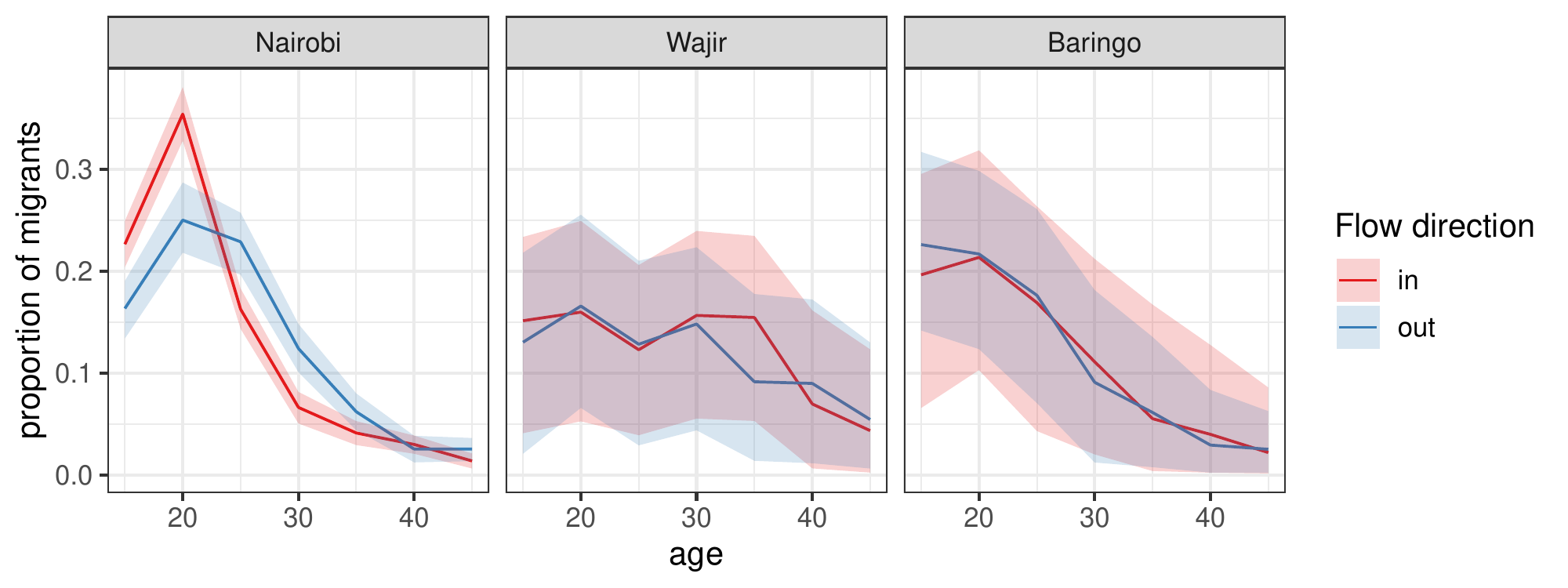}
\caption{Estimates of age distribution of in- and out-migration. \label{mig_A_both}}
\end{subfigure}
\caption{Estimates and 95\% credible intervals of migration components for three counties. \label{fig_mig}}
\end{figure}

In addition to mortality, there is substantial variation in patterns in
migration across Kenyan counties. Figure \ref{fig_mig} shows estimates of all migration components in the three case
study counties. For total in-migration and out-migration estimates (Figure \ref{mig_both}), flows into and out of Nairobi are much larger, with net
in migration reaching almost 400,000 people per year. Flows into Wajir
are much smaller (\textless10,000 people), and in 2019 Baringo had net
out-migration of around 10,000. The estimated age patterns of migration for the three counties are
also shown in Figure \ref{mig_A_both}. Some differences exist, with Nairobi's
immigrants much more concentrated around age 20.

\newpage
\subsection{Age-time multiplier}\label{multiplier}

\begin{figure}[h!]
\centering
\includegraphics{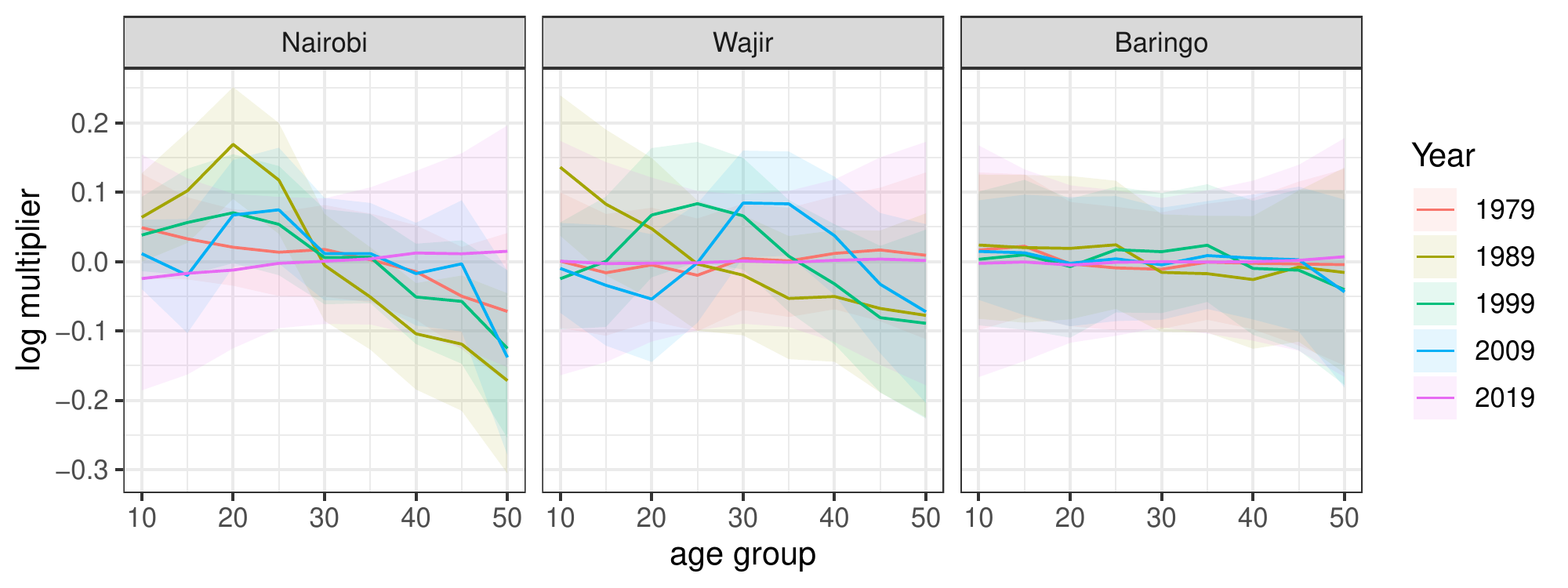}
\caption{Age-time specific multipliers for three counties. \label{eps}}
\end{figure}

Figure \ref{eps} shows the age-time multipliers $\varepsilon$ for the three example counties. For Baringo, the multipliers are essentially always zero on the log scale. This observation is true for the majority of counties (see Appendix \ref{appendix_results} for plots for additional counties), which suggests that most of the patterns over age and time are captured well by the mortality and migration components. For county-years where multipliers do deviate from zero, estimates are at most around 10\% of the total population magnitude, and usually between 0-0.05\%. For example for Nairobi, the estimated multiplier suggests that, after accounting for the expected mortality and migration components, in 1989, we see an additional increase around age 20 (of around 10\%) and an additional decrease of around 10\% at the oldest age group. 

\clearpage

\hypertarget{model-evaluation}{%
\subsection{Model evaluation}\label{model-evaluation}}

A national census was run in Kenya in 2019. While the microlevel data
are not yet publicly available (for example, via IPUMS), the resulting
population counts by age, sex and county have been published by the
Kenya National Bureau of Statistics (Kenya National Bureau of Statistics
2019). We can therefore evaluate the 2019 projections from our model
with the actual counts from the 2019 census.

We extracted census population counts by age, sex and county from a PDF
file containing the results following code provided by Alexander (2020).
We compared the 2019 projections from the Bayesian cohort component
projection model with these counts and calculated several summary metrics. We define the relative error $e_g$ for a particular group \(g\) as 
\begin{equation}
e_g = \frac{y_{g,2019} - \hat{\eta}_{g, 2019}}{y_{g,2019}},
\end{equation}
where $y_{g,2019}$ refers to the census-based population count for that population and $\hat{\eta}_{g, 2019}$ to the model-based projection. A group $g$ can refer to an age-county or age-district group, for example. 

Based on the errors, we calculate mean, median, and root mean squared errors by age group and for the total population. We compared
 these results to the results of a similar linear extrapolation model,
 where the population in 2019 was estimated based on applying the same
 proportion change seen between the 1999-2009 censuses.  Errors are summarized over districts, as estimates by county are not possible with the linear extrapolation method (as we only have one previous set of census observations by county).

% Firstly, the root mean squared
% error: \begin{equation}
% RMSE = \sqrt{\frac{\sum_g \left(\frac{ \hat{p}_{g, 2019} - p^{\text{census}}_{g 2019}}{p^{\text{census}}_{g 2019}}\right)^2}{G}}
% \end{equation}

% where \(\hat{p}_{g, 2019}\) is the estimated population in a particular
% group \(g\), \(p^{\text{census}}_{g 2019}\) is the equivalent population
% from the census and \(N\) is the size of the group. Here, the \(g\) can
% refer to any combination of age group, and district/county. We also calculated the mean and median bias of estimates, where bias is calculated as
% \begin{equation}
% \text{bias}_g = \log p^{\text{census}}_{g 2019} - \log \hat{p}_{g, 2019}
% \end{equation}
% We compared
% these results to the results of a similar linear extrapolation model,
% where the population in 2019 was estimated based on applying the same
% proportion change seen between the 1999-2009 censuses.

Error summaries by age group and for the total population are shown in Table \ref{mse}. In general, the
Bayesian model projections are within \textasciitilde1\% of the census
populations. The magnitudes of the RMSEs for the simple linear
interpolation is 3-10 times higher than that of the Bayes CCP. The bias results suggest that the point estimate from the Bayes CCP is often slightly lower than the census observation, whereas linear interpolation substantially over-estimates population counts. 

\begin{table}[h!]
\centering
\begin{tabular}{r|rr|rr|rr}
\toprule
\multicolumn{1}{l|}{}                 & \multicolumn{2}{c|}{Mean error}                                    & \multicolumn{2}{c|}{Median error}                                  & \multicolumn{2}{c}{RMSE}                                         \\ \hline
Age group                             & \multicolumn{1}{r}{Interpolation} & \multicolumn{1}{r|}{Bayes CCP} & \multicolumn{1}{r}{Interpolation} & \multicolumn{1}{r|}{Bayes CCP} & \multicolumn{1}{r}{Interpolation} & \multicolumn{1}{r}{Bayes CCP} \\ \hline
15                                    & -0.070                            & -0.058                         & 0.128                             & 0.010                          & 0.013                             & 0.005                         \\
20                                    & -0.228                            & -0.081                         & -0.040                            & -0.033                         & 0.016                             & 0.005                         \\
25                                    & -0.266                            & -0.019                         & 0.018                             & 0.006                          & 0.020                             & 0.005                         \\
30                                    & -0.146                            & 0.043                          & 0.035                             & 0.047                          & 0.020                             & 0.005                         \\
35                                    & -0.254                            & -0.161                         & -0.008                            & -0.179                         & 0.035                             & 0.012                         \\
40                                    & -0.058                            & -0.074                         & 0.185                             & -0.048                         & 0.038                             & 0.009                         \\
45                                    & -0.246                            & -0.065                         & 0.049                             & 0.011                          & 0.061                             & 0.021                         \\ \hline
\multicolumn{1}{l|}{Total population} & \multicolumn{1}{r}{-0.101}        & \multicolumn{1}{r|}{-0.045}    & \multicolumn{1}{r}{0.119}         & \multicolumn{1}{r|}{0.011}     & 0.031                             & 0.010                         \\ \bottomrule
\end{tabular}
\caption{Summary of errors in district population sizes by age group comparing 2019 census counts with two methods, linear interpolation and the Bayesian cohort component projection model (Bayes CPP).}
\label{mse}
\end{table}

We also calculated the coverage of the 95\% prediction intervals of the Bayesian cohort component projection model estimates for 2019, compared to the observed 2019 census counts, and the proportion of census counts above and below the prediction intervals. If the model is well-calibrated, on average around 90\% of the observed census counts should fall within the 90\% prediction intervals, and 5\% of observation should fall above and below the interval. Table \ref{coverage} reports coverage by age group, and suggests that in general the coverage of the credible intervals matches expectations. However, in some age groups, there is a relative bias towards observations falling below the interval rather than above.

\begin{table}[h!]
\centering
\begin{tabular}{rrrr}
\toprule
Age Group & Prop in interval & Prop above & Prop below \\ \hline
15        & 0.89             & 0.02       & 0.08       \\
20        & 0.89             & 0.00          & 0.09       \\
25        & 0.89             & 0.04       & 0.04       \\
30        & 0.91             & 0.06       & 0.02       \\
35        & 0.87             & 0.01       & 0.09       \\
40        & 0.92             & 0.04       & 0.04      \\
45        & 0.87             & 0.04       & 0.05       \\ \bottomrule
\end{tabular}
\caption{Proportion of 2019 census county counts falling within, above, and below the 90\% prediction intervals as estimated by the Bayesian CPP model.}
\label{coverage}
\end{table}

We also calculated the probability integral transform (PIT) to assess the consistency between the 2019 projections and observed counts. Results are presented in Appendix \ref{PIT}.

\hypertarget{discussion}{%
\section{Discussion}\label{discussion}}

In this paper we proposed a Bayesian cohort component projection
framework to estimate adult subnational populations with limited amounts
of data available. The model uses information on population and
migration counts from censuses, as well as mortality patterns from
national schedules, to reconstruct populations based on cohorts moving
through time. The modeling framework also naturally extends to allow
projection of populations. In addition, the model ensures the national
populations implied by the sum of subnational areas agree with national
pre-published UN WPP estimates.

The model was used to estimate and project populations of women of
reproductive ages (WRA) for counties in Kenya over the period 1979-2019.
Results suggested continued growth of WRA populations in all districts,
and accelerated growth in particular in areas such as Nairobi and
Northeastern. The mortality component of the modeling framework
highlighted the stagnating progress through the 1990s and 2000s, largely
due to HIV/AIDS, but more recent mortality declines. The estimates from
the Kenyan example also highlighted substantial differences in internal
migration patterns across the nation.

The model requires only inputs from national censuses and WPP estimates,
which are available for the majority of countries. Thus, while the model
was tested on estimation in Kenya, the methodology is applicable to a
wide range of countries with very little alterations. For example, there
is currently census microdata available for almost 100 counties on the
IPUMS-International website.

Based on a series of validation measures, the proposed model outperformed a benchmark model of linear interpolation. In addition to having lower performance than Bayes CCP, note that with the
simple interpolation method, it is not possible to get estimates by county easily,
because 2009 is the first year which the counties as they are today were
recorded. In
addition, another advantage of the Bayesian model is that the
population estimates also have an associated uncertainty level, and
that estimating not only population counts but also mortality and
migration rates allows us to better understand the drivers of population
change by county.

There are several other advantages and contributions of this modeling
framework to the estimation of subnational populations. The model is
governed by a cohort component projection model, tracking cohorts as
they move through time. This has advantages over more aggregate
techniques such as interpolation and extrapolation, because it allows us to understand trends in overall population as a process governed by separate components that add or remove population. In addition, this process takes
into account intercensal events such as trends in HIV/AIDS mortality and produces estimates and projections with uncertainty. 

Secondly, the modeling framework proposes a parsimonious model for
internal net-migration across subnational areas. In cohort component
models, it is often the case that migration components are assumed to be
negligible or considered to just be the residual once mortality has been
taken into account. Very little data usually exists on migration
patterns, and estimation of all migration components by age, region and
year becomes very intensive. After observing key patterns in the data,
we proposed a net-migration model which separates migration patterns
into independent age and time components. The result is an age-specific
net migration model with parameters that are easier to estimate when
data are limited.

More broadly, one of the contributions of our proposed framework over
existing work in this area is the use of mortality and migration models
that have relatively strong functional forms, which allow plausible
estimates to be produced even in the absence of good-quality data. Our
approach to modeling mortality through the use of characteristic age
patterns is inspired by the long demographic tradition of using model
life tables where information on mortality are sparse.

While we have illustrated the utility of this approach in data-limited
contexts, the framework can naturally be extended to include additional
sources of data. For example, if there exist observations of
age-specific mortality rates at the subnational level (even at some
ages), these data could be used as inputs to the mortality model.
If more reliable data on internal migration flows were available, the
existing migration process model --- which assumes a fixed age schedule
with varying magnitude over time --- could be reformulated to be more
flexible. In general, to be able to handle population projection in a low data availability context, the model proposed here includes mortality and mortality process models that separate age- and time-trends into independent effects. Additional age-time specific effects were then captured by the multiplier $\varepsilon$. If more data are available, the underlying process models could be extended to better understand these age-time specific effects and how they relate to either mortality or migration. 

Another possible extension of this framework is to include other total
population estimates such as those from WorldPop as additional ``data''
that could be used to inform estimates. As such, we view this
methodology and subnational population estimates produced from it as
complementary to estimates produced by other efforts such as the
WorldPop project. As mentioned in Section 2, the primary goal of the
WorldPop estimates is to produce extremely fine-grained estimates of
total population, whereas we are more interested in understanding
population patterns by age and sex and the underlying components of
population change within larger subnational areas.

The incorporation of a cohort component projection model into a
probabilistic setting allows for different sources of uncertainty, such
as sampling and non-sampling error, to be included into the modeling
process. The Bayesian hierarchical framework allows information from
different data sources to be consolidated without the need for
post-estimation redistribution changes as is often the case with
subnational population estimation (Swanson and Tayman 2012). In
addition, it allows for increased flexibility in modeling population
processes compared to traditional deterministic techniques, while still
keeping the basis of an underlying demographic process.

\newpage
\appendix

\hypertarget{other-potential-data-sources-1}{%
\section{\texorpdfstring{Other potential data sources
\label{other_data}}{Other potential data sources }}\label{other-potential-data-sources-1}}

We use census data and WPP estimates as inputs to the model. There are
other available data sources that could be used as inputs. These sources
and the reasons for not including them are discussed below.

\hypertarget{mortality-1}{%
\subsection{Mortality}\label{mortality-1}}

Mortality is estimated at the subnational level based on national
patterns of mortality from WPP, as well as changes in subnational
population counts over time. Thus, no explicit information on
subnational mortality levels is used; mortality is estimated based on
likely patterns at the national level and intercensal changes in
population. There are two main sources for subnational mortality data in
Kenya that are not included as data inputs.

Firstly, the Demographic and Health Survey (DHS) collects information
about sibling mortality histories. Adult mortality can be calculated from these data using the
sibling history method, where cohorts of siblings are constructed and
age-specific mortality rates are calculated based on when they died.
Previous research has illustrated sibling data produces relatively
reliable estimates at the national level (Masquelier, 2013). However, the DHS does not ask the location of residents of
the siblings who died, thus the data cannot be used to inform differentials in subnational mortality. 

A second source of information on subnational mortality comes from a
question about household deaths, that was collected in the most recent
census (2009). This can be used to obtain death probabilities by age.
However, previous research has found that the value of \(_{45}q_{15}\)
implied by household deaths is often much lower or higher than other
mortality sources (Masquelier et al. 2017). Indeed, mortality
information from census household deaths is excluded from other
mortality analyses due to its unreliable nature (e.g.~child mortality,
see UN-IGME (2017)). As such, we chose to omit this information for now.
Future work will investigate this data source to see if it can be used
to inform age patterns of mortality by subnational region.

\hypertarget{migration-1}{%
\subsection{Migration}\label{migration-1}}

There are two other potential sources of information on internal
migration in Kenya that are not included as data inputs. Firstly, the
census also includes a question about how many years the person has
resided in their current locality of residence, referring to the
district level. The question is asked in the 1999 and 2009 censuses. Based on the year of the census and the age of the
respondent, as well as how many years they indicated they had lived in
the current locality, the implied year and age of in-migration can be
calculated. However, this method gave much lower numbers of in-migration
compared to those implied by the `location one year ago' question. As
such this information was not used in the model.

Secondly, the DHS contains some information about
migration.\footnote{Note that questions about migration in the DHS differ by country. The migration questions in the Kenya DHS are quite minimal; however for other countries there may be more useful data available.}
For Kenya, it is possible to obtain information about the proportion of
the population who moved to a particular province in the year before the
survey. However, when compared to corresponding data from the census,
there were large discrepancies, and trends in DHS proportions were
erratic over time.

\newpage 

\hypertarget{app-model}{%
\section{\texorpdfstring{Full Model Specification
\label{appendix_model}}{Full Model Specification }}\label{appendix_model}}

The full model specification is described below. 

\subsection{Population}

\subsubsection{Cohort component projection model}

The underlying population by age group, year and county \(\eta_{a,t,c}\) is
\begin{equation}
\eta_{a,t,c}= \left(\eta_{a-1,t-1,c} \cdot (1- \gamma_{a-1,t-1,c})\right) \cdot (1 + \phi_{a-1,t-1,c}) \cdot(\varepsilon_{a-1,t-1,c}),
\end{equation} where \(\gamma_{a,t,c}\) is the conditional probability
of death in age group \(a\), year \(t\) and county \(c\),
\(\phi_{a,t,c}\) is net migration (that is, in- minus out-migration) as
a proportion of population size and \(\varepsilon_{a,t,c}\) is an
additional age-year-county multiplier.

\subsubsection{Data model}

The data model is: \begin{eqnarray}
    \log y_i|\eta_{a,t,c} &\sim& \begin{cases}
    N\left(\log \eta_{a[i],t[i],c[i]}, s^2_{y}[i]\right) \text{ if } t= 2009,\\
    N\left(\log \sum_{c \in d[i]}(\eta_{a[i],t[i],c[i]}), s^2_{y}[i]\right) \text{ if } t<2009,
    \end{cases}
\end{eqnarray} where  \(y_i\) is \(i\)th observed population count, \(s^2_y\) is the sampling error based on the fact
that the micro-data in IPUMS is a 10\% sample. The second case of the
above equation dictates that if we have observations prior to 2009, we
can only relate these to \(\eta_{a,t,c}\)'s that have been summed to the
district level.

\subsubsection{National constraints}
We constrain the sum of the county populations by age and year to be within approximately 10\% of the national estimates produced by WPP:
\begin{eqnarray}
 \Lambda_{a,t} < &\sum_{c}\log \eta_{a,t,c}& \leq \Omega_{a,t},\\
    \log \Lambda_{a,t} &\sim& N(\log 0.9WPP_{a,t}, 0.1^2)T(, \log WPP_{a,t}),\\
    \log \Omega_{a,t} &\sim& N(\log 1.1WPP_{a,t}, 0.1^2)T(\log WPP_{a,t},).
\end{eqnarray}
\subsubsection{Priors on first year and age group}

The cohort component projection framework requires priors to be placed on populations in the first
year and age group. We use the following priors: \begin{eqnarray}
    \log \eta_{1,t,c} &\sim& N(\log WPP_{1,t} + \log \text{prop}_{1,t,c}, 0.01^2), \\
    \log \eta_{a,1,c} &\sim& N(\log WPP_{a,1} + \log \text{prop}_{a,1,c}, 0.01^2),
\end{eqnarray} where \(WPP_{a,t}\) is the national-level population
count from WPP in the relevant age group and year, and
\(\text{prop}_{a,t,c}\) is the proportion of the total population in the
relevant age, year and county, which was calculated based on
interpolating census year proportions and assuming the proportion of a
district's population in each county was constant at a level equal to
2009.

\subsection{Mortality}
The model for mortality is
as \begin{eqnarray}
    \text{logit} \gamma_{a,t,c} &=& \alpha_{0,c} + Y_{a,0} + \beta_{t,c,1}\cdot Y_{a,1} + \beta_{t,c,2}\cdot Y_{a,2},
\end{eqnarray} where \(Y_{a,0}\) is the mean age-specific logit
mortality schedule of the national mortality curves and \(Y_{,1}\) and
\(Y_{,2}\) are the first two principal components derived from
national-level mortality schedules. Modeling on the logit scale ensures
the death probabilities are between zero and one.

The county-specific mortality intercepts are modeled using a Normal
distribution centered at zero:

\[
\alpha_{0,c}| \sigma^2_{\alpha} \sim N(0, \sigma^2_{\alpha}),
\]

The county-specific coefficients \(\beta_{t,c, k}\) are modeled as
fluctuations around a national mean: \begin{eqnarray}
    \beta_{t,c, k} &=& B^{nat}_{t,k} + \delta_{t,c, k},\\
    \delta_{t,c,k}|\delta_{t-1,c, k},\sigma^2_{\delta} &\sim& N(\delta_{t-1,c, k},\sigma^2_{\delta}),
\end{eqnarray} where \(B^{nat}_{a,t,k}\) are the national coefficient on
principal components, derived from WPP data. The county-specific
fluctuations are modeled as a random walk.

\subsection{Migration}

\subsubsection{Process model}
The process model for the net-migration is:
\begin{eqnarray}
    \phi_{a,t,c} &=& \frac{\psi_{a,t,c}^{in} - \psi_{a,t,c}^{out}}{\eta_{a-1,t-1,c}},\\
    \psi_{a,t,c}^{in} &=& \Psi_{t,c}^{in}\cdot \Pi^{in}_{a,c},\\
    \psi_{a,t,c}^{out} &=& \Psi_{t,c}^{out}\cdot \Pi^{out}_{a,c},
\end{eqnarray} where \(\Psi_{t,c}^{in}\) and \(\Psi_{t,c}^{out}\) are the total number
of in- and out-migrants, respectively, and \(\Pi^{in}_{a,c}\) and
\(\Pi^{out}_{a,c}\) are the relevant age distributions. We model the total counts as a second order random walk to
impose a certain level of smoothness in the counts over time:
\begin{eqnarray}
    \Psi_{1,c}^{in} &\sim& U(0, y_c),\\
    \log \Psi_{2,c}^{in}|\Psi_{1,c}^{in}, \sigma^2_{in} &\sim& N(\log \Psi_{1,c}^{in}, \sigma^2_{in}),\\
    \log \Psi_{t,c^{in}}|\Psi_{(t-2):(t-2),c}^{in}, \sigma^2_{in}  &\sim& N(2 \log \Psi_{t-1,c}^{in}- \log \Psi_{t-2,c}^{in}, \sigma^2_{in}),\\
    \Psi_{1,c}^{out} &\sim& U(0, y_k[c]),\\
    \log \Psi_{2,c}^{out}|\Psi_{1,c}^{out}, \sigma^2_{out} &\sim& N(\log \Psi_{1,c}^{out}, \sigma^2_{out}),\\
    \log \Psi_{t,c}^{out}|\Psi_{(t-2):(t-2),c}^{out}, \sigma^2_{out}  &\sim& N(2 \log \Psi_{t-1,c}^{out} -  \log \Psi_{t-2,c}^{out}, \sigma^2_{out}).
\end{eqnarray}

where $y_c$ refers to the observed total population for county $c$ based on the census in the first observation period. 

We place Uniform priors on the non-normalized age distributions of in- and
out-migration, with equal prior probability on each age group:
\begin{eqnarray}
    \Pi^{in*}_{a,c} &\sim& \text{Uniform}(0,1),\\
    \Pi^{out*}_{a,c} &\sim& \text{Uniform}(0,1).
\end{eqnarray} 

We then normalize the age distributions as
\begin{eqnarray}
    \Pi^{in}_{a,c} &=& \frac{\Pi^{in*}_{a,c}}{\sum_a \Pi^{in*}_{a,c}},\\
\Pi^{out}_{a,c} &=& \frac{\Pi^{out*}_{a,c}}{\sum_a \Pi^{out*}_{a,c}}.
\end{eqnarray} 

\subsubsection{Data model}
We relate the observed age-specific in- and out-migration
counts in the censuses, denoted \(M_i^{in}\) and \(M_i^{out}\), respectively, to
the underlying true counts \(\psi_{a,t,c}^{in}\) and \(\psi_{a,t,c}^{out}\) through
the following data model: \begin{eqnarray}
    \log M_i^{in}|\psi_{a,t,c}^{in} &\sim& \begin{cases}
    N\left(\log \psi^{in}_{a[i],t[i],c[i]}, s^2_{in}[i]\right) \text{ if } t[i]= 2009,\\
    N\left(\log \sum_{c \in d[i]}(\psi^{in}_{a[i], t[i],c[i]}), s^2_{in}[i]\right) \text{ if } t[i]<2009,
    \end{cases}\\
       \log M_i^{out}|\psi_{a,t,c}^{out} &\sim& \begin{cases}
    N\left(\log \psi_{a[i],t[i],c[i]}^{out}, s^2_{out}[i]\right) \text{ if } t[i]= 2009,\\
    N\left(\log \sum_{c \in d[i]}(\psi_{a[i],t[i],c[i]}^{out}), s^2_{out}[i]\right) \text{ if } t[i]<2009.
    \end{cases}
\end{eqnarray} 

\subsubsection{Constraint}
Using the fact that the sum of all internal migration for a particular age group and year should be around zero, we implement the following constraint: 
\begin{eqnarray}
 \sum_{c} -0.1 \eta_{a,t,c} < &\sum_{c}\psi^{in}_{a,t,c} - \sum_{c}\psi^{out}_{a,t,c} & \leq \sum_{c}0.1 \eta_{a,t,c}.
\end{eqnarray} 
The constrain states that the difference between the sum of all in- and
out-migration flows across all counties cannot be more than \(\pm 10\%\)
of the total estimated national population for that particular age group
and year.

\subsection{Age-time multiplier}
We model multipliers
on the log scale, and to ensure identifiability we assume the mean of
the sum of the log multipliers is zero. This constraint is implemented
through the re-parameterization: \begin{eqnarray}
\log \bm{\varepsilon}_{1:A,t,c} &=& \mathbf{D(DD')}^{-1} \bm{\zeta}_{1:(A-1),t,c}, \\
\bm{\zeta}_{a,t,c} &\sim& N(0, \sigma^2_{\zeta}),
\end{eqnarray} where \(\mathbf D\) is first-order difference matrix (with
\({D}_{i,i} = -1\), \({D}_{i,i+1}=1\), and \({D}_{i,j} = 0\) otherwise) such that $\zeta_{a,t,c} = \log {\varepsilon}_{a,t,c} - \log \bm{\varepsilon}_{a-1,t,c}$.

\hypertarget{priors-on-variance-parameters}{%
\subsection{Priors on variance
parameters}\label{priors-on-variance-parameters}}

All variance parameters that are estimated
(\(\sigma^2_{\alpha}, \sigma^2_{\delta}, \sigma^2_{\Psi}, \sigma^2_{in}   \text{ and } \sigma^2_{{out}}\))
have half-Normal standard priors placed on them, i.e. \[
\sigma \sim N^{+}(0,1).
\]

\newpage

\hypertarget{age-patterns-in-migration-data}{%
\section{\texorpdfstring{Age patterns in migration data
\label{mig_data}}{Age patterns in migration data }}\label{age-patterns-in-migration-data}}

\begin{figure}[h!]
\includegraphics{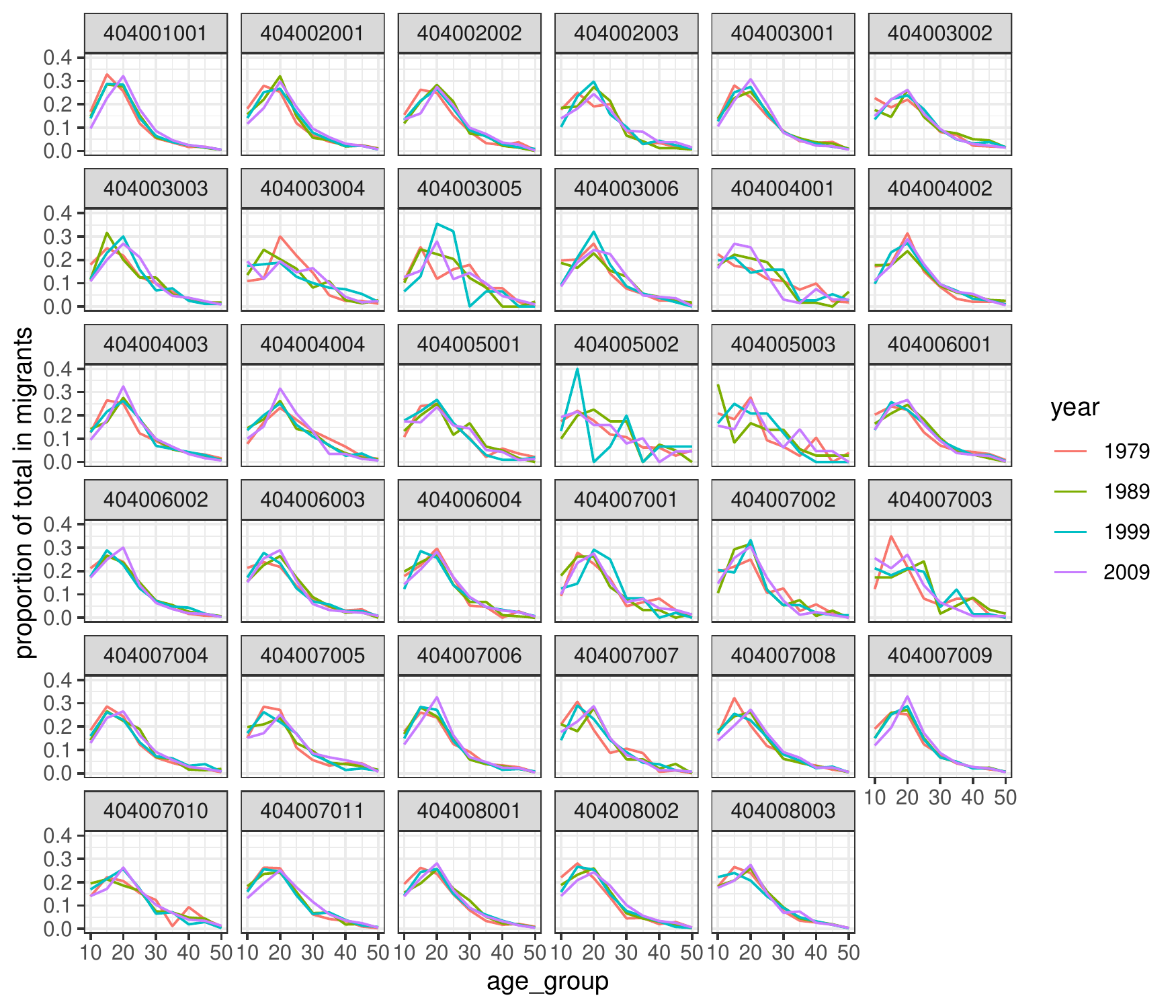}
\caption {Observed age patterns of in-migration from Kenyan censuses, 1979-2009.}
\label{mig_census_in}
\end{figure}

\begin{figure}[h!]
\includegraphics{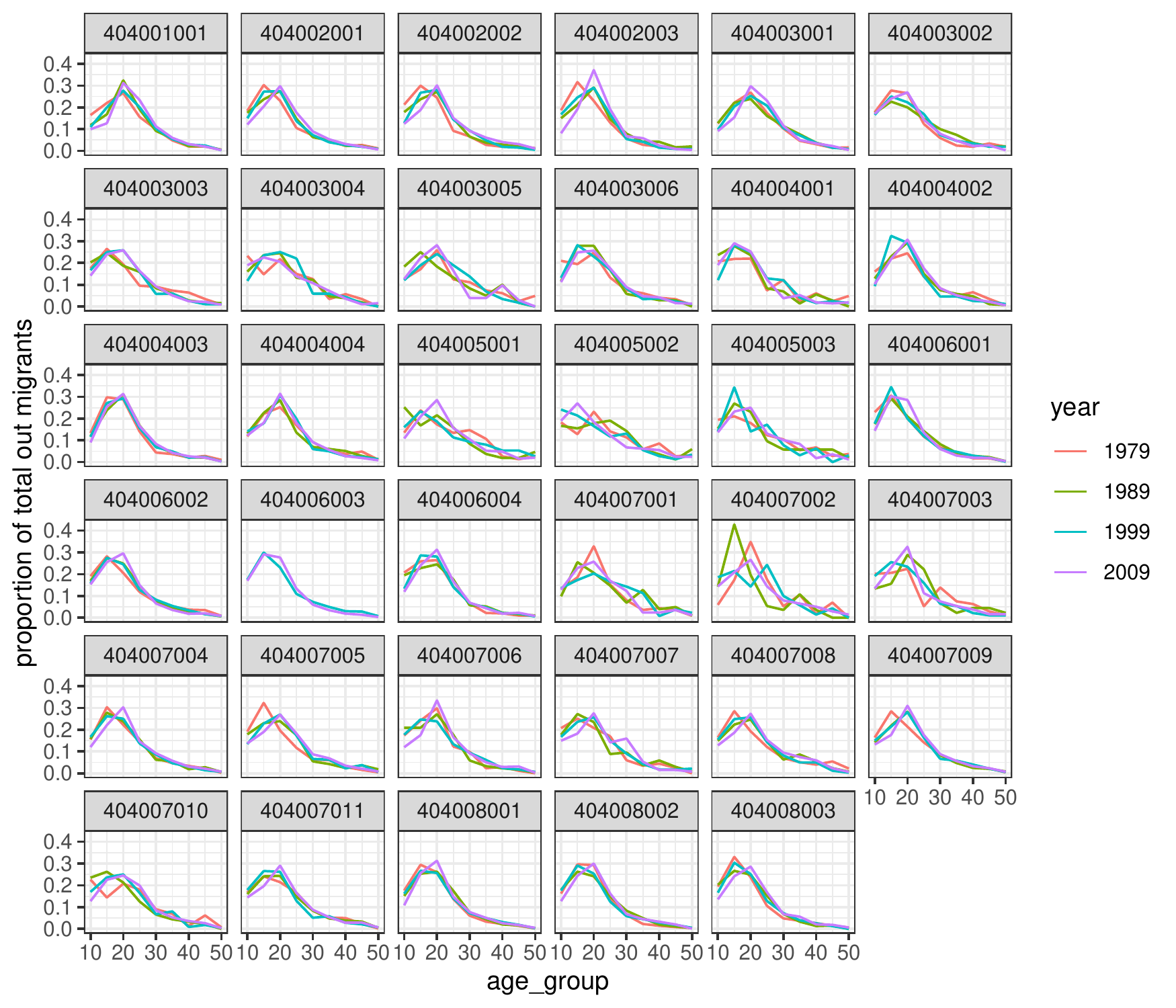}
\caption {Observed age patterns of out-migration from Kenyan censuses, 1979-2009.}
\label{mig_census_out}
\end{figure}

In the Bayesian cohort component model, specifically in the migration process model, we assume the age distribution of in- and out-migrants by count is constant over time (see Equations 13 and 14). This is a somewhat strong assumption and was made to ensure identifiability of all parameters in the model in cases where we do not have very much data. While the assumption is relatively strong, it was motivated by age patterns observed in census data. Figures \ref{mig_census_in} and \ref{mig_census_out} show the proportion of all in- and out-migrants by age group for each year and district, and illustrate that the age patterns remain remarkably constant over time. For reference, the broad areas covered by the IPUMS districts are listed in Table \ref{table_district}.


\begin{table}[h!]
\footnotesize
\begin{tabular}{l|l}
\toprule
District  & Areas                                                                                                                                                                                            \\ \hline
404001001 & Nairobi East, Nairobi North, Nairobi West, Westlands                                                                                                                                             \\
404002001 & Gatanga, Gatundu, Githunguri, Kiambu (Kiambaa), Kikuyu, 
Lari, Muranga, Nyandarua, Ruiru,   Thika, Maragua \\
404004001 & Chalbi, Laisamis, Marsabit, Moyale                                                                                                                                                               \\
404004002 & Garba Tulla, Igembe, Imenti, Isiolo,   Maara, Meru, Tharaka, Tigania, Meru                                                \\
404004003 & Embu, Kangundo, Kibwezi, Machakos, Makueni, Mbeere, Mbooni, Mwala, Nzaui,   Yatta                                                                                                                \\
404004004 & Kitui North, Kitui South (Mutomo), Kyuso, Mwingi                                                                                                                                                 \\
404005001 & Fafi, Garissa, Ijara, Lagdera                                                                                                                                                                    \\
404005002 & Wajir East, Wajir North, Wajir South, Wajir West                                                                                                                                                 \\
404005003 & Mandera Central, Mandera East, Mandera West                                                                                                                                                      \\
404006001 & Bondo, Rarieda, Siaya                                                                                                                                                                            \\
404006002 & Kisumu East, Kisumu West, Nyando                                                                                                                                                                 \\
404006003 & Homa Bay, Kuria East, Kuria West, Migori, Rachuonyo, Rongo, Suba                                                                                                                                 \\
404002002 & Nyeri North, Nyeri South                                                                                                                                                                         \\
404006004 & Borabu, Gucha, Gucha South, Kisii Central, Kisii South, Manga, Masaba,   Nyamira, North Kisii                                                                                                    \\
404007001 & Turkana Central, Turkana North, Turkana South                                                                                                                                                    \\
404007002 & Pokot Central, Pokot North, West Pokot                                                                                                                                                           \\
404007003 & Samburu Central, Samburu East, Samburu North                                                                                                                                                     \\
404007004 & Kwanza, Trans Nzoia East, Trans Nzoia West                                                                                                                                                       \\
404007005 & Baringo, Baringo North, East Pokot, Koibatek, Laikipia East, Laikipia   North, Laikipia West                                                                                                     \\
404007006 & Eldoret East, Eldoret West, Wareng, Uasin Gishu                                                                                                                                                  \\
404007007 & Keiyo, Marakwet, Elgeyo Markwet                                                                                                                                                                  \\
404007008 & Nandi Central, Nandi East, Nandi North, Nandi South, Tinderet                                                                                                                                    \\
404007009 & Kaijiado Central, Kaijiado North, Loitoktok, Molo, Naivasha, Nakuru,   Nakuru North, Kajiado                                                                                                     \\
404002003 & Kirinyaga                                                                                                                                                                                        \\
404007010 & Narok North, Narok South, Trans Mara                                                                                                                                                             \\
404007011 & Bomet, Buret, Kericho, Kipkelion, Sotik                                                                                                                                                          \\
404008001 & Butere, Emuhaya, Hamisi, Kakamega, Lugari, Mumias, Vihiga, Butere/Mumias                                                                \\
404008002 & Bungoma East, Bungoma North, Bungoma South, Bungoma West, Mt. Elgon                                                                                                                              \\
404008003 & Bunyala, Busia, Samia, Teso North, Teso South                                                                                                                                                    \\
404888001 & Waterbodies                                                                                                                                                                                      \\
404003001 & Kilindini, Kilindini, Mombasa                                                                                                                                                                    \\
404003002 & Kinango, Kwale, Msambweni                                                                                                                                                                        \\
404003003 & Kaloleni, Kilifi, Malindi                                                                                                                                                                        \\
404003004 & Tana Delta, Tana River                                                                                                                                                                           \\
404003005 & Lamu                                                                                                                                                                                             \\
404003006 & Taita, Taveta, Taita Taveta                                                                                                                                                                      \\ \bottomrule
\end{tabular}
\caption{IPUMS district codes and areas covered}
\label{table_district}
\end{table}

\clearpage 

\hypertarget{app-results}{%
\section{\texorpdfstring{Additional results
\label{appendix_results}}{Additional results}}\label{appendix_results}}
In this section we highlight several other components that are estimated within the model; specifically the coefficients on the first and second principal components. Results are illustrated on three example counties: Nairobi, Wajir and Baringo. Additionally, we show estimates for the age-time multiplier for all counties. 

\begin{figure}[h!]
\centering
\begin{subfigure}[b]{\textwidth}
\includegraphics{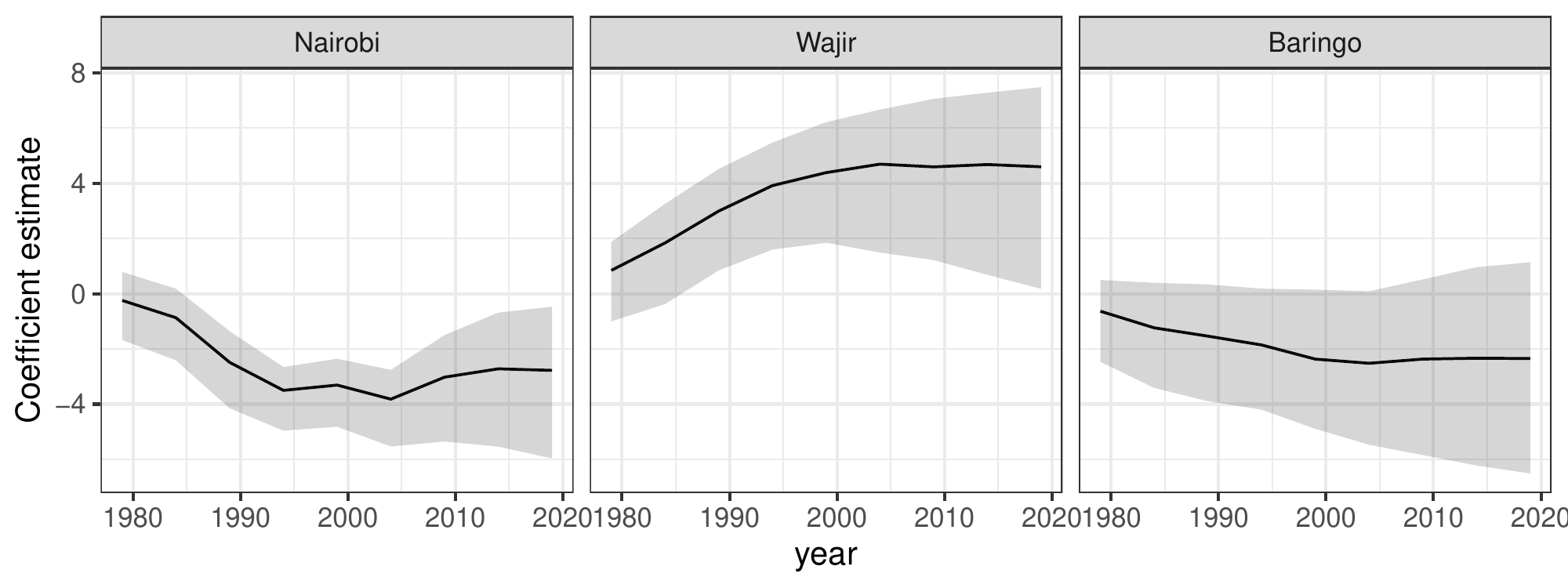}
\caption{Estimates for first component (region deviations from national mortality trends)}
\end{subfigure}
\hfill
\begin{subfigure}[b]{\textwidth}
\includegraphics{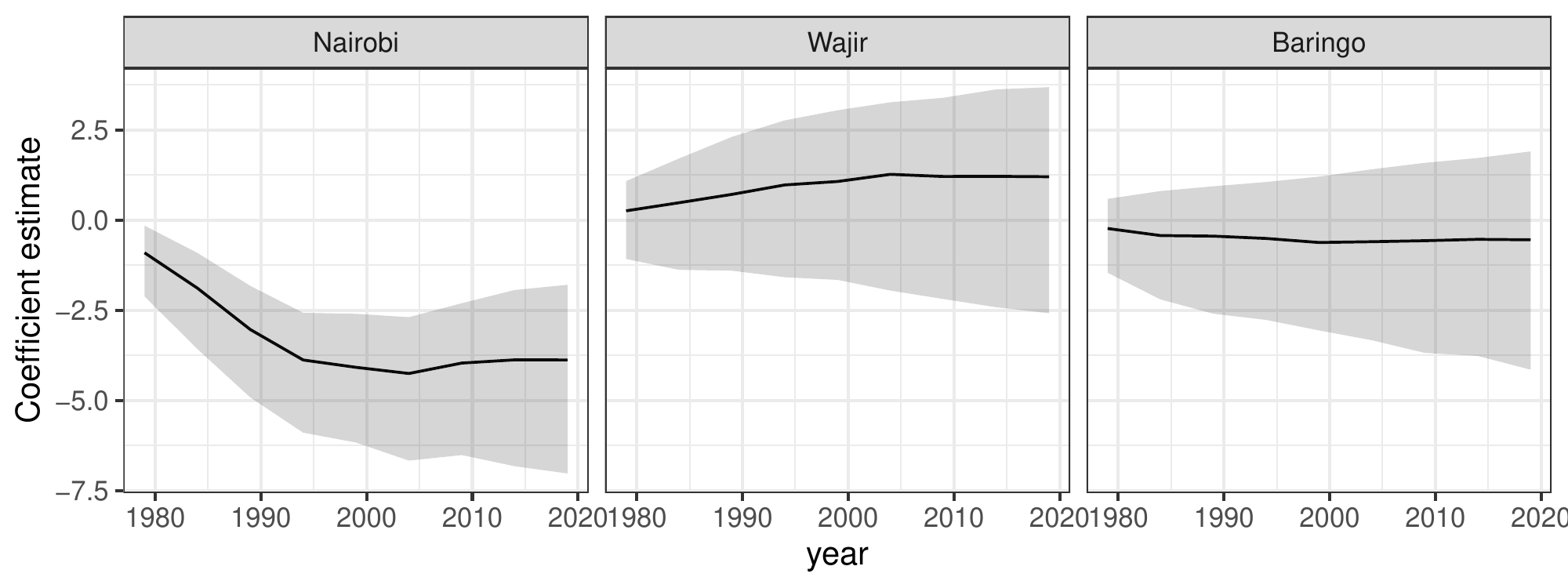}
\caption{Estimates for second component (region deviations from national HIV/AIDS mortality)}
\end{subfigure}
\caption{County-specific deviations from national-level mortality improvements (first component) and HIV/AIDS mortality (second
component) for three counties. \label{delta2}}
\end{figure}

Figure \ref{delta2} shows estimates over time of the coefficient of the first and second principal component within the mortality model (i.e. $\beta_{tc,1}$ and $\beta_{tc,2}$). Broadly, the first principal component relates to overall mortality improvement, and the second relates to the effect of the HIV/AIDS epidemic. Coefficients on the first component suggest mortality improvement is relatively slow in Nairobi, and better than the national average in Wajir. Based on patterns on the second principal component, there is evidence to suggest that the effect of HIV/AIDS epidemic was relatively small in Wajir (Figure \ref{delta2}). In both cases, estimates for Baringo are not significantly different from the national mean. 

\begin{figure}[h!]
\centering
\includegraphics[width = \textwidth]{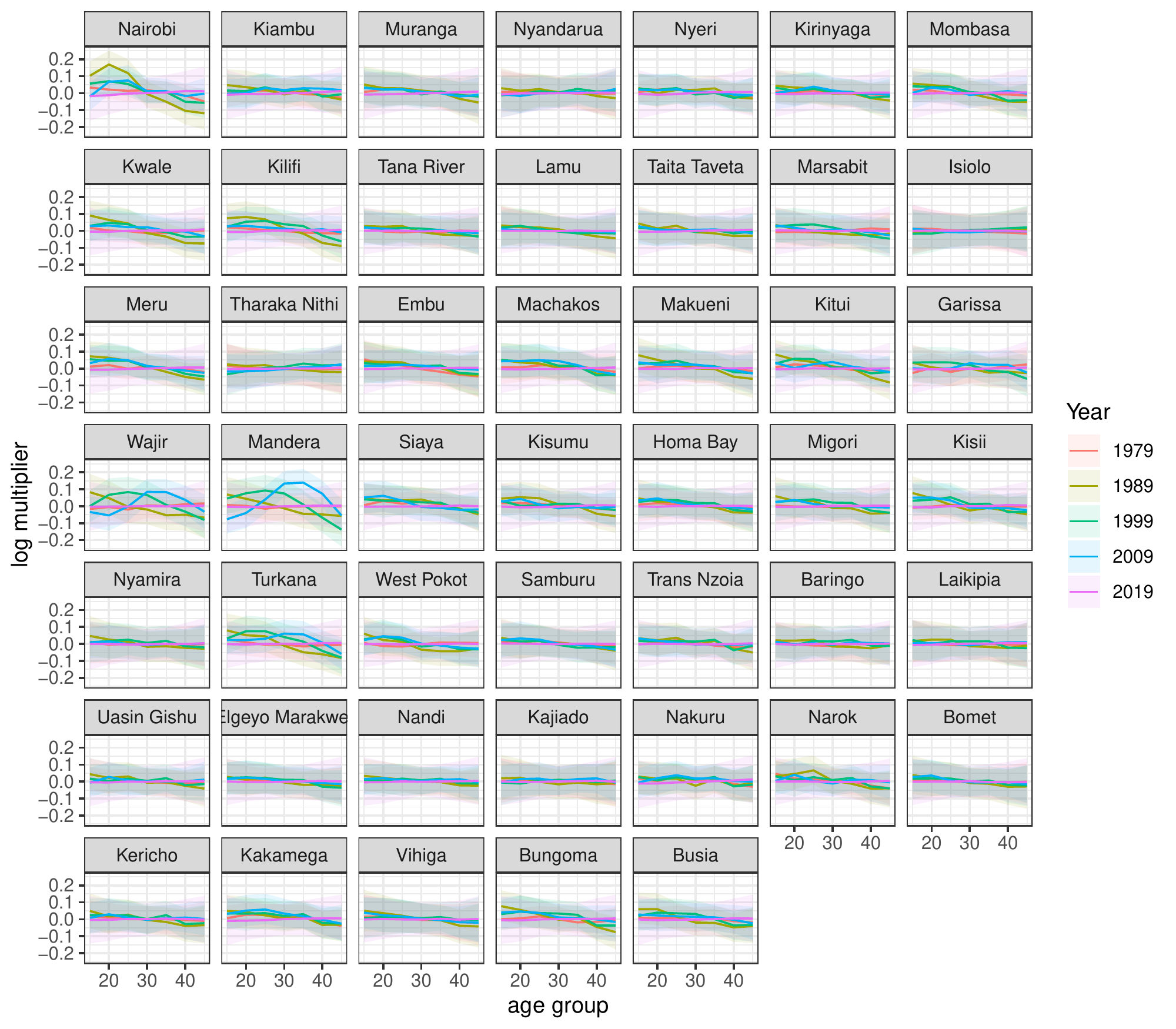}
\caption{Age-time specific multipliers for all counties.}
\label{fit_eps_all}
\end{figure}

Figure \ref{fit_eps_all} shows the estimated age-time specific multiplier for all counties. As can be seen, the estimates on the log scale are very close to zero for the majority of age groups, years and counties. 

\clearpage
\hypertarget{pit-histogram}{%
\section{\texorpdfstring{PIT histogram
\label{PIT}}{PIT histogram. }}\label{histogram}}

A Probability Integral Transform (PIT) histogram is a tool for evaluating the similarity between model projections and left out observations. The predictive distributions of the projections are compared with the actual observations. 

For each observation $j$ for 2019 (i.e. each population count by age group and county) we have observation $y_j$ from the 2019 census, and sample $\hat{\eta}_j^{(S)}$ from the corresponding posterior distribution (with a total of $S$ samples). The PIT for observation $j$ was calculated as

\begin{equation}
PIT_j = \frac{\sum_{s = 1}^S \hat{\eta}_{j}^{(s)}\leq y_j}{S}.
\end{equation}

\begin{figure}[h!]
\centering
\includegraphics[width = 0.7\textwidth]{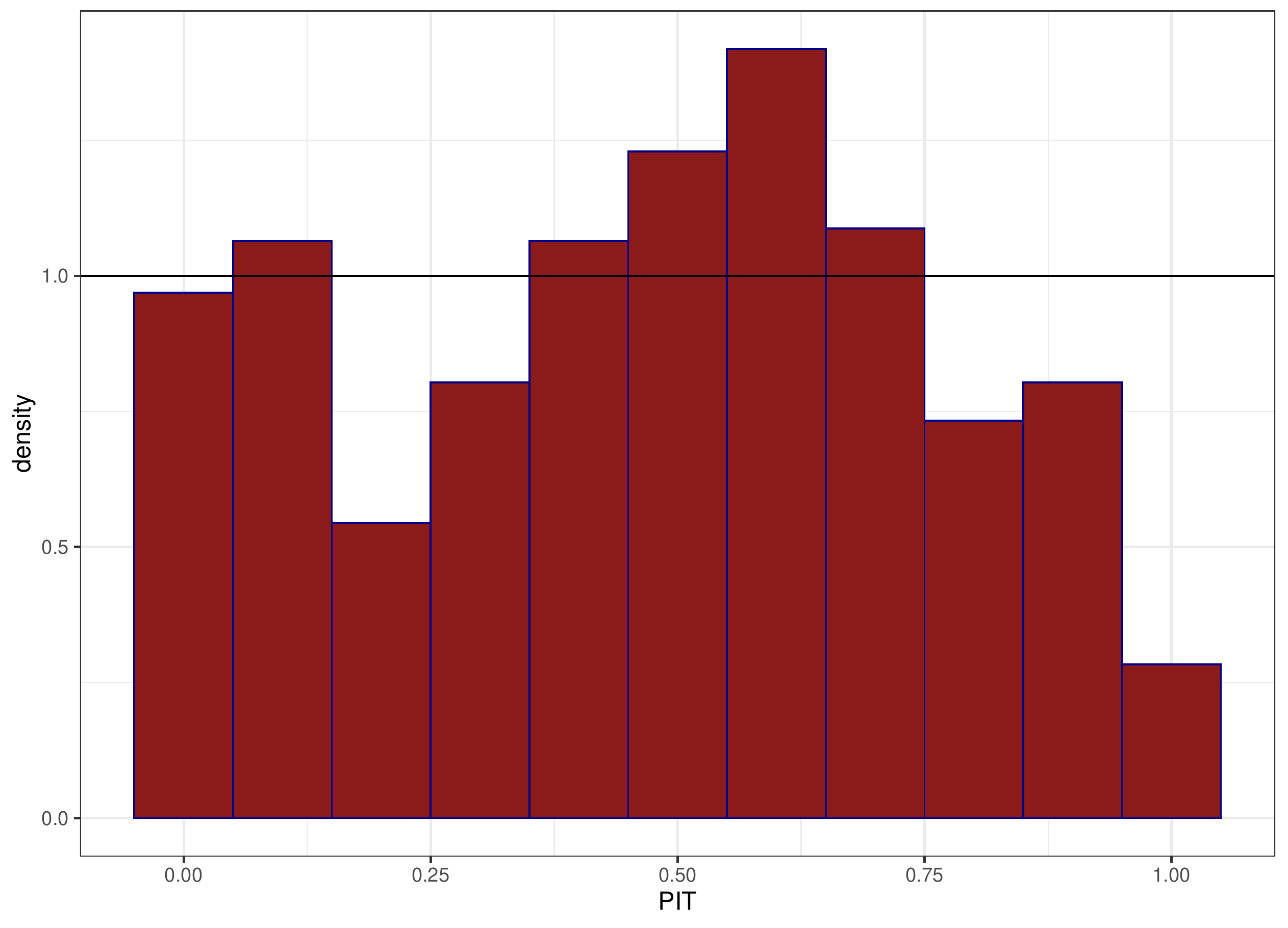}
\caption{PIT histogram comparing projected 2019 population counts with observed 2019 census counts.}
\label{fig_pit}
\end{figure}

If the predictive distribution is well calibrated, the result should be a uniform distribution of PIT values. Figure \ref{fig_pit} shows the PIT histogram for 2019. The relatively high density in the middle of the distribution suggests the model is somewhat over-dispersed, and the low density towards 1 suggests the upper bound of population projections is in general too conservative. 

\newpage

\hypertarget{references}{%
\section*{References}\label{references}}
\addcontentsline{toc}{section}{References}

\hypertarget{refs}{}
\leavevmode\hypertarget{ref-alexander2018}{}%
Alexander, Monica, and Leontine Alkema. 2018. ``Global Estimation of
Neonatal Mortality Using a Bayesian Hierarchical Splines Regression
Model.'' \emph{Demographic Research} 38: 335--72.

\leavevmode\hypertarget{ref-alexander2017}{}%
Alexander, Monica, Emilio Zagheni, and Magali Barbieri. 2017. ``A
Flexible Bayesian Model for Estimating Subnational Mortality.''
\emph{Demography} 54 (6): 2025--41.

\leavevmode\hypertarget{ref-alexanderclean}{}%
Alexander, Rohan. 2020. ``Telling Stories with Data: Extracting Data
from PDFs.'' Available at:
\url{https://www.tellingstorieswithdata.com/03-03-pdfs.html}.

\leavevmode\hypertarget{ref-alkemau52014}{}%
Alkema, Leontine, and Jin Rou New. 2014. ``Global Estimation of Child
Mortality Using a Bayesian B-Spline Bias-Reduction Model.'' \emph{The
Annals of Applied Statistics} 8 (4): 2122--49.

\leavevmode\hypertarget{ref-alkema2011fertility}{}%
Alkema, Leontine, Adrian E Raftery, Patrick Gerland, Samuel J Clark,
François Pelletier, Thomas Buettner, and Gerhard K Heilig. 2011.
``Probabilistic Projections of the Total Fertility Rate for All
Countries.'' \emph{Demography} 48 (3): 815--39.

\leavevmode\hypertarget{ref-bijak2008bayesian}{}%
Bijak, Jakub. 2008. ``Bayesian Methods in International Migration
Forecasting.'' \emph{International Migration in Europe: Data, Models and
Estimates}, 255--88.

\leavevmode\hypertarget{ref-bijak2016}{}%
Bijak, Jakub, and John Bryant. 2016. ``Bayesian Demography 250 Years
After Bayes.'' \emph{Population Studies} 70 (1): 1--19.
\url{https://doi.org/10.1080/00324728.2015.1122826}.

\leavevmode\hypertarget{ref-bryant2013}{}%
Bryant, John R, and Patrick J Graham. 2013. ``Bayesian Demographic
Accounts: Subnational Population Estimation Using Multiple Data
Sources.'' \emph{Bayesian Analysis} 8 (3): 591--622.

\leavevmode\hypertarget{ref-bryant2018bayesian}{}%
Bryant, John, and Junni L Zhang. 2018. \emph{Bayesian Demographic
Estimation and Forecasting}. CRC Press.

\leavevmode\hypertarget{ref-clark2016}{}%
Clark, Samuel J. 2016. ``A General Age-Specific Mortality Model with an
Example Indexed by Child or Child/Adult Mortality.'' \emph{arXiv
Preprint arXiv:1612.01408}.

\leavevmode\hypertarget{ref-congdon1997}{}%
Congdon, P, S Shouls, and S Curtis. 1997. ``A Multi-Level Perspective on
Small-Area Health and Mortality: A Case Study of England and Wales.''
\emph{Population, Space and Place} 3 (3): 243--63.

\leavevmode\hypertarget{ref-ihme2017u5}{}%
GBD 2016 Mortality Collaborators (IHME). 2017. ``Global, Regional, and
National Under-5 Mortality, Adult Mortality, Age-Specific Mortality, and
Life Expectancy, 1970-2016: A Systematic Analysis for the Global Burden
of Disease Study 2016.'' \emph{The Lancet} 390 (10100): 1084--1150.

\leavevmode\hypertarget{ref-gelman2020}{}%
Gelman, Andrew, Aki Vehtari, Daniel Simpson, Charles C. Margossian, Bob Carpenter, Yuling Yao, Lauren Kennedy, Jonah Gabry, Paul-Christian Bürkner, and Martin Modrák 2020. ``Bayesian workflow.'' \emph{arXiv preprint arXiv:2011.01808}.

\leavevmode\hypertarget{ref-girosi2008}{}%
Girosi, Federico, and Gary King. 2008. \emph{Demographic Forecasting}.
Princeton University Press.

\leavevmode\hypertarget{ref-he2017china}{}%
He, Chunhua, Li Liu, Yue Chu, Jamie Perin, Li Dai, Xiaohong Li, Lei
Miao, et al. 2017. ``National and Subnational All-Cause and
Cause-Specific Child Mortality in China, 1996-2015: A Systematic
Analysis with Implications for the Sustainable Development Goals.''
\emph{The Lancet Global Health} 5 (2): e186--e197.

\leavevmode\hypertarget{ref-ipumsgeo}{}%
IPUMS. 2018.
\url{https://international.ipums.org/international-action/variables/GEO2_KE\#description_section}.

\leavevmode\hypertarget{ref-dhs2015}{}%
Kenya National Bureau of Statistics. 2015. ``Kenya Demographic and
Health Survey 2014.'' Rockville, MD, USA.
\url{http://dhsprogram.com/pubs/pdf/FR308/FR308.pdf}.

\leavevmode\hypertarget{ref-kenya2019census}{}%
---------. 2019. ``2019 Kenya Population and Housing census Volume I:
Population by County and Sub-County.'' Available at:
\url{https://www.knbs.or.ke/?wpdmpro=2019-kenya-population-and-housing-census-volume-i-population-by-county-and-sub-county}.

\leavevmode\hypertarget{ref-leasure2020national}{}%
Leasure, Douglas R, Warren C Jochem, Eric M Weber, Vincent Seaman, and
Andrew J Tatem. 2020. ``National Population Mapping from Sparse Survey
Data: A Hierarchical Bayesian Modeling Framework to Account for
Uncertainty.'' \emph{Proceedings of the National Academy of Sciences}
117 (39): 24173--9.

\leavevmode\hypertarget{ref-census2017methods}{}%
Leddy, Robert M. 2017. ``Methods for Calculating 5-Year Age Group
Population Estimates by Sex for Subnational Areas.'' Available at:
\url{https://www2.census.gov/programs-surveys/international-programs/about/global-mapping/subntl-pop-est-methods-pgs-uscb-dec16.pdf}.

\leavevmode\hypertarget{ref-leecarter1992}{}%
Lee, Ronald D., and Lawrence R. Carter. 1992. ``Modeling and Forecasting
U.s. Mortality.'' \emph{Journal of the American Statistical Association}
87 (419): 659--71. \url{http://www.jstor.org/stable/2290201}.

\leavevmode\hypertarget{ref-leslie1945}{}%
Leslie, Patrick H. 1945. ``On the Use of Matrices in Certain Population
Mathematics.'' \emph{Biometrika} 33 (3): 183--212.

\leavevmode\hypertarget{ref-lim2016gbd}{}%
Lim, Stephen S, Nancy Fullman, Christopher JL Murray, and Amanda Jayne
Mason-Jones. 2016. ``Measuring the Health-Related Sustainable
Development Goals in 188 Countries:: A Baseline Analysis from the Global
Burden of Disease Study 2015.'' \emph{The Lancet}, 1--38.

\leavevmode\hypertarget{ref-linard2012population}{}%
Linard, Catherine, Marius Gilbert, Robert W Snow, Abdisalan M Noor, and
Andrew J Tatem. 2012. ``Population Distribution, Settlement Patterns and
Accessibility Across Africa in 2010.'' \emph{PloS One} 7 (2): e31743.

\leavevmode\hypertarget{ref-masquelier2013}{}%
Masquelier, B., 2013. ``Adult mortality from sibling survival data: a reappraisal of selection biases.'' \emph{Demography}, 50(1), pp.207-228.

\leavevmode\hypertarget{ref-masquelier2017}{}%
Masquelier, Bruno, Jeffrey W Eaton, Patrick Gerland, François Pelletier,
and Kennedy K Mutai. 2017. ``Age Patterns and Sex Ratios of Adult
Mortality in Countries with High Hiv Prevalence.'' \emph{AIDS} 31:
S77--S85.

\leavevmode\hypertarget{ref-mpc}{}%
Minnesota Population Center. 2017. ``Integrated Public Use Microdata
Series, International: Version 6.5 {[}Dataset{]}.'' Available at:
\url{https://international.ipums.org/international/}.

\leavevmode\hypertarget{ref-new2017levels}{}%
New, Jin Rou, Niamh Cahill, John Stover, Yogender Pal Gupta, and
Leontine Alkema. 2017. ``Levels and Trends in Contraceptive Prevalence,
Unmet Need, and Demand for Family Planning for 29 States and Union
Territories in India: A Modelling Study Using the Family Planning
Estimation Tool.'' \emph{The Lancet Global Health} 5 (3): e350--e358.

\leavevmode\hypertarget{ref-plummer2003}{}%
Plummer, Martyn. 2003. ``JAGS: A Program for Analysis of Bayesian
Graphical Models Using Gibbs Sampling.'' In \emph{Proceedings of the 3rd
International Workshop on Distributed Statistical Computing}. Vienna,
Austria.

\leavevmode\hypertarget{ref-raftery2012}{}%
Raftery, Adrian E, Nan Li, Hana {\v{S}}ev{\v{c}}{\'\i}kov{\'a}, Patrick Gerland, and Gerhard
K Heilig. 2012. ``Bayesian Probabilistic Population Projections for All
Countries.'' \emph{Proceedings of the National Academy of Sciences} 109
(35): 13915--21.

\leavevmode\hypertarget{ref-schmertmann2018}{}%
Schmertmann, Carl, and Marcos Roberto Gonzaga. 2018. ``Bayesian
Estimation of Age-Specific Mortality and Life Expectancy for Small Areas
with Defective Vital Records.''

\leavevmode\hypertarget{ref-schmertmann2013}{}%
Schmertmann, Carl P, Suzana M Cavenaghi, Renato M Assunção, and Joseph E
Potter. 2013. ``Bayes Plus Brass: Estimating Total Fertility for Many
Small Areas from Sparse census Data.'' \emph{Population Studies} 67 (3):
255--73.

\leavevmode\hypertarget{ref-schmertmann2014}{}%
Schmertmann, Carl, Emilio Zagheni, Joshua R Goldstein, and Mikko
Myrskylä. 2014. ``Bayesian Forecasting of Cohort Fertility.''
\emph{Journal of the American Statistical Association} 109 (506):
500--513.

\leavevmode\hypertarget{ref-sevcikova2017}{}%
Sevcikova, Hana, Adrian E Raftery, and Patrick Gerland. 2018.
``Probabilistic Projection of Subnational Total Fertility Rates.''
\emph{Demographic Research} 38(60): 1843-1884.

\leavevmode\hypertarget{ref-swanson2012subnational}{}%
Swanson, David A, and Jeff Tayman. 2012. \emph{Subnational Population
Estimates}. Vol. 31. Springer Science \& Business Media.

\leavevmode\hypertarget{ref-tatem2013millennium}{}%
Tatem, Andrew J, Andres J Garcia, Robert W Snow, Abdisalan M Noor,
Andrea E Gaughan, Marius Gilbert, and Catherine Linard. 2013.
``Millennium Development Health Metrics: Where Do Africa's Children and
Women of Childbearing Age Live?'' \emph{Population Health Metrics} 11
(1): 11.

\leavevmode\hypertarget{ref-igme2017}{}%
UN-IGME. 2017. ``Levels and Trends in Child Mortality: Report 2017.''
Available at:
\url{http://www.childmortality.org/files_v21/download/IGME\%20report\%202017\%20child\%20mortality\%20final.pdf}.

\leavevmode\hypertarget{ref-unpd2019}{}%
UNPD. 2019a. ``World Population Prospects: The 2019 Edition.'' Available
at: \url{http://esa.un.org/wpp/}.

\leavevmode\hypertarget{ref-unpd2019method}{}%
---------. 2019b. ``World Population Prospects: The 2019 Edition.
Methodology of the United Nations Population Estimates and
Projections.'' Available at:
\url{https://esa.un.org/unpd/wpp/publications/Files/WPP2019_Methodology.pdf}.

\leavevmode\hypertarget{ref-census2017}{}%
U.S. census Bureau. 2017. ``Subnational Population by Sex, Age, and
Geographic Area.'' Available at:
\url{https://www.census.gov/geographies/mapping-files/time-series/demo/international-programs/subnationalpopulation.html}.

\leavevmode\hypertarget{ref-wachter2014}{}%
Wachter, Kenneth W. 2014. \emph{Essential Demographic Methods}. Harvard
University Press.

\leavevmode\hypertarget{ref-wakefield2019estimating}{}%
Wakefield, Jon, Geir-Arne Fuglstad, Andrea Riebler, Jessica Godwin,
Katie Wilson, and Samuel J Clark. 2019. ``Estimating Under-Five
Mortality in Space and Time in a Developing World Context.''
\emph{Statistical Methods in Medical Research} 28 (9): 2614--34.

\leavevmode\hypertarget{ref-wakefield2019estimating}{}%
Wardrop, N. A., W. C. Jochem, T. J. Bird, H. R. Chamberlain, D. Clarke, D. Kerr, L. Bengtsson, S. Juran, V. Seaman, and A. J. Tatem.``Spatially disaggregated population estimates in the absence of national population and housing census data.'' \emph{Proceedings of the National Academy of Sciences} 115(14): 3529-3537.

\leavevmode\hypertarget{ref-westoff2006}{}%
Westoff, Charles F, and Anne R Cross. 2006. ``The Stall in the Fertility
Transition in Kenya.''

\leavevmode\hypertarget{ref-wheldon2013}{}%
Wheldon, Mark C, Adrian E Raftery, Samuel J Clark, and Patrick Gerland.
2013. ``Reconstructing Past Populations with Uncertainty from
Fragmentary Data.'' \emph{Journal of the American Statistical
Association} 108 (501): 96--110.

\leavevmode\hypertarget{ref-wheldon2016}{}%
---------. 2016. ``Bayesian Population Reconstruction of Female
Populations for Less Developed and More Developed Countries.''
\emph{Population Studies} 70 (1): 21--37.

\leavevmode\hypertarget{ref-who2016}{}%
World Health Organization (WHO). 2016a. ``WHO Methods and Data Sources
for Life Tables 1990-2015.'' Available at:
\url{http://www.who.int/healthinfo/statistics/LT_method.pdf}.

\leavevmode\hypertarget{ref-who2016monitor}{}%
---------. 2016b. \emph{World Health Statistics 2016: Monitoring Health
for the Sdgs Sustainable Development Goals}. World Health Organization.

\leavevmode\hypertarget{ref-worldpop}{}%
WorldPop. 2018. ``Population Movements: Mapping Population Mobility and
Connectivity.'' \url{www.worldpop.org}.

\end{document}